\documentstyle[eqsecnum,preprint,aps,epsf]{revtex}
\tightenlines

\newcommand{\reseteqnum}{\setcounter{equation}{0}}
\newcommand{\nn}{\nonumber}
\def\lsim{\raise0.3ex\hbox{$<$\kern-0.75em\raise-1.1ex\hbox{$\sim$}}}
\def\gsim{\raise0.3ex\hbox{$>$\kern-0.75em\raise-1.1ex\hbox{$\sim$}}}

\newcommand{\ovl}[1]{\overline{#1}}
\newcommand{\wt}[1]{\widetilde{#1}}
\newcommand{\eqn}[1]{(\ref{#1})}
\newcommand{\p}{\partial}
\newcommand{\bpsi}{{\overline{\psi}}}
\newcommand{\bq}{{\overline{q}}}
\newcommand{\tr}{{\rm tr}}
\newcommand{\bra}[1]{\left\langle #1\right|}
\newcommand{\ket}[1]{\left| #1\right\rangle}
\newcommand{\braket}[2]{\vev{#1 | #2}}
\newcommand{\vev}[1]{\left\langle #1 \right\rangle}
\newcommand{\VEV}[3]{\left\langle #1\left| #2 \right| #3\right\rangle}
\begin{document}

%\draft

\title{
\vspace{-3.0cm}
\begin{flushright}
{\normalsize hep-lat/0109022}\\
{\normalsize UTHEP-447}\\
{\normalsize UTCCP-P-112}\\
\end{flushright}
Chiral properties of domain-wall fermions\\
from eigenvalues of 4 dimensional Wilson-Dirac operator}

\author{S.~Aoki and Y.~Taniguchi}
\address{
  Institute of Physics, University of Tsukuba,
  Tsukuba, Ibaraki 305-8571, Japan\\
}

\date{\today}

\maketitle

\begin{abstract}
We investigate chiral properties of the domain-wall fermion (DWF)
system by using the four-dimensional hermitian Wilson-Dirac operator.
We first derive a formula which connects a chiral symmetry
breaking term in the five dimensional DWF Ward-Takahashi identity with
the four dimensional Wilson-Dirac operator,
and simplify the formula in terms of only the eigenvalues of the operator,
using an ansatz for the form of the eigenvectors.
For a given distribution of the eigenvalues,
we then discuss the behavior of the chiral symmetry breaking term 
as a function of the fifth dimensional length.
We finally argue the chiral property of the DWF formulation in the limit of the
infinite fifth dimensional length, in
connection with spectra of the hermitian Wilson-Dirac operator 
in the infinite volume limit as well as in the finite volume.

\end{abstract}

%11.15Ha : Lattice gauge theory (see also 12.38.G Lattice QCD calculations)
%12.38Gc : Lattice QCD calculations (see also 11.15.H Lattice gauge theory)

%\narrowtext

%%%%%%%%%%%%%%%%%%%%%%%%%%%%%Section 1%%%%%%%%%%%%%%%%%%%%%%%%%%%%%%%
\section{Introduction}

A suitable definition of the chiral symmetry has been a long standing 
problem in lattice field theories. Recently an ultimate solution to this 
problem seems to appear in the form of the Ginsberg-Wilson relation
\cite{GW,Luscher98}.
Two explicit examples of the lattice fermion operators which satisfies the 
Ginsberg-Wilson relation have been found so far:
One is the perfect lattice Dirac operator constructed via the renormalization
group transformation\cite{perfect1,perfect2} and the other is the
overlap Dirac(OD) operator\cite{Neuberger98,KN99}
derived from the overlap formalism\cite{NN94}
or from the domain-wall fermion(DWF)\cite{Kaplan92,Shamir93,Shamir95} in
the limit of the infinite length of the 5th dimension.
Since the explicit form is simpler for the latter, a lot of numerical
investigations\cite{Blum-Soni,AIKT,Blum98,cppacs-dwf,RBC,practical,EHN98,MILC}
as well as analytic considerations\cite{HJL98,Kikukawa99}
have been carried out
for the domain-wall fermion or the overlap Dirac fermion.

Recent numerical investigations for the domain-wall
fermion\cite{cppacs-dwf,RBC}, however, bring puzzling results, which are
summarized as follows.

Analytic considerations suggest that the overlap or domain-wall fermion
works well at sufficiently weak gauge coupling or equivalently for
sufficiently smooth gauge configurations\cite{HJL98,Kikukawa99}.
Initial numerical investigations supported this
result\cite{Blum-Soni,AIKT,Blum98}.
On the other hand, further investigations indicate that the domain-wall
fermion at stronger coupling ceases to describe the massless fermion 
even in the $N_5 \rightarrow \infty$ limit\cite{cppacs-dwf,RBC}, 
where $N_5$ is the number of sites in the 5th dimension. 
Analytic results\cite{IN,BS,GS} in the strong coupling limit are
controversial. However the latest one\cite{BBS} also suggests 
that the DWF does not work in the limit.

It has been argued\cite{cppacs-dwf} that this result may be understood 
by the relation between the phase structure of the lattice QCD with the 
4 dimensional Wilson fermion\cite{Aoki-phase} and zero eigenvalues
of the 4 dimensional Wilson-Dirac operator. It is well-known that
the zero eigenvalues cause a trouble for the domain-wall fermion
(or the overlap)\cite{Shamir95,HJL98,Kikukawa99},
therefore it is natural to consider that the success/failure of DWF
depends on the absence/presence of the zero eigenvalues.
On the other hand, the zero eigenvalues of the 4 dimensional Wilson
Dirac operator, denoted as $D_W$, is related to the parity-flavor
breaking order parameter, $\vev{\bar q i\gamma_5\tau^3 q}$, where
$\tau^3$ is the 2 $\times$ 2 flavor matrix. Introducing
the external source $H$ coupled to $\bar q i\gamma_5\tau^3 q$,
the following relation is easily derived.
\begin{eqnarray}
\vev{\bar q i\gamma_5\tau^3 q} &=& -\lim_{H\rightarrow 0^+}
\lim_{V\rightarrow \infty}
\frac{1}{V}{\rm Tr}\frac{i\gamma_5\tau^3}{D_W + i\gamma_5\tau^3 H}
\nn\\&=&
-\lim_{H\rightarrow 0^+}\lim_{V\rightarrow \infty}\frac{1}{V}{\rm tr}
\left[\frac{i\gamma_5}{D_W + i\gamma_5 H}-
\frac{i\gamma_5}{D_W - i\gamma_5 H}
\right] \nonumber \\
&=& -i \lim_{H\rightarrow 0^+}\lim_{V\rightarrow \infty}\frac{1}{V}{\rm tr}
\left[\frac{1}{H_W + i H}-\frac{1}{H_W - i H}
\right] \nn\\
&=& -i \lim_{H\rightarrow 0^+}\lim_{V\rightarrow \infty}\frac{1}{V}
\sum_n \VEV{\lambda_n}{\frac{1}{\lambda_n + i H}-\frac{1}{\lambda_n - i H}}
{\lambda_n} \nonumber \\
&=& -i \lim_{H\rightarrow 0^+}
\int d\ \lambda \rho_{H_W}
(\lambda)\left[\frac{1}{\lambda + i H}-\frac{1}{\lambda - i H}\right]\nn\\
&=& -i\int d\ \lambda \rho_{H_W}(\lambda) (-2\pi i)\delta(\lambda)
= -2\pi \rho_{H_W}(0),
\end{eqnarray}
where $\lambda_n$ and $\vert \lambda_n\rangle$ are the
eigenvalue and the eigenstate of the hermitian Wilson-Dirac
operator $H_W =\gamma_5 D_W$ and $\rho_{H_W}(\lambda)$ is the
density of eigenvalues of $H_W$, defined by
\begin{equation}
\rho_{H_W}(\lambda) =\lim_{V\rightarrow \infty}
\frac{1}{V} \sum_n \delta(\lambda-\lambda_n) .
\end{equation}

\begin{figure}[htb]
\begin{center}
  \leavevmode
  \epsfxsize=11cm \epsfbox{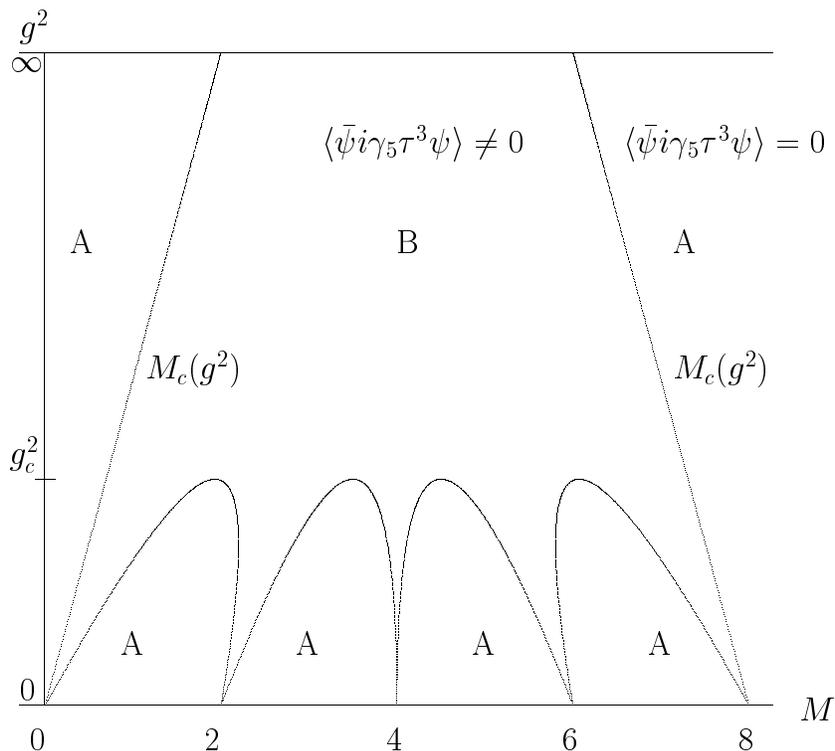}
\caption{The expected phase structure of lattice QCD with the Wilson fermion
in the $M$--$g^2$ plane.
In the region A, $\langle \bar\psi i\gamma_5\tau^3\psi\rangle =0$ 
(the parity-flavor symmetric phase), while
$\langle \bar\psi i\gamma_5\tau^3\psi \rangle \not=0$
(the parity-flavor broken phase) in the region B.
}
\label{fig:phase}
\end{center}
\end{figure}

The expected phase structure of lattice QCD with the Wilson fermion
\cite{Aoki-phase} is given in Fig.~\ref{fig:phase}, where $g$
is the gauge coupling and $M$ is the mass parameter.
In the region B, $\vev{\bar q i\gamma_5\tau^3 q}\not= 0$,
thus the density of the zero eigenvalues is nonzero.
According to this phase structure, the DWF is successful
between $ 0 < M < 2$ in the weak coupling limit of QCD.
This allowed region of $M$ agrees with the analytic consideration of 
DWF\cite{Shamir93}.
Once the gauge coupling becomes nonzero, the allowed region of 
$M$ shrinks and moves to larger values. This property has been predicted
by the mean-field analysis\cite{AT} and has numerically been 
observed\cite{cppacs-dwf}.
If the coupling becomes large so that $\beta=6/g^2 < \beta_c$,
the allowed range of $M$ disappears and the massless fermion 
ceases to exist in the domain-wall QCD (DWQCD).

Numerical results mentioned before\cite{cppacs-dwf,RBC} 
that DWF does not work in the strong coupling region
seem to agree with the above expectation.
This interpretation, however, has difficulties.
Let us summarize the recent numerical result\cite{cppacs-dwf}, where
the quenched DWQCD has been investigated with the RG improved gauge 
action as well as
the ordinary plaquette gauge action. With the former gauge action
the quenched DWQCD works at $\beta = 2.6$ and fails at $\beta=2.2$, 
suggesting that
$2.2 < \beta_c < 2.6$, while with the latter gauge action 
the quenched DWQCD fails even 
at $\beta = 6.0$, indicating $\beta_c > 6.0$.
One problem is that the latter condition $\beta_c > 6.0$ with the
plaquette gauge contradicts the previous numerical investigation\cite{AKU},
which concludes that the region without parity-flavor breaking
(: the allowed region in the present case) exists $\beta = 6.0$.
A more serious problem is that the numerical analysis for the
density of eigenvalues indicates non-zero value of $\rho_{H_W}(0)$ 
at any $\beta$\cite{EHN99} with the plaquette action, and the similar 
conclusion is obtained by the same analysis with the RG improved action
\cite{cppacs-nagai}.
If this is true, the phase structure in Fig.\ref{fig:phase} is incorrect
in the quenched QCD:
the gap of the parity-flavor breaking phase never shows up  in the
weak coupling region, leading to the conclusion that DWF does not work 
at all in the quenched QCD.

This complicated situation is summarized as follows. 
While being consistent with the
numerical result of the DWQCD for the plaquette action\cite{cppacs-dwf,RBC},
the numerical analysis for the density of the
eigenvalues\cite{EHN99,cppacs-nagai} contradicts the phase structure of
QCD with Wilson fermion\cite{Aoki-phase,AKU}
and the numerical result of DWQCD with the RG action\cite{cppacs-dwf}. 
Furthermore the numerical result of the DWQCD with the plaquette action
seems apparently inconsistent with the numerical result for the phase
structure of the Wilson fermion with the same gauge action.

In this paper, we try to resolve the above mutual inconsistency,
by analysing eigenvalues of the 4 dimensional hermitian Wilson-Dirac operator.
In Sec.~\ref{sec:action} the action and the axial Ward-Takahashi
identity are given for the latter use.
In Sec.~\ref{sec:effective}, 
we derive the formula for the explicit chiral symmetry breaking term of
DWQCD, $m_{5q}$, 
in terms of eigenvalues and eigenstates of the modified hermitian
Wilson-Dirac operator. 
In Sec.~\ref{sec:expansion}, assuming the form of the eigenstates,
the formula derived in Sec.~\ref{sec:effective}
is reduced to a much simpler expression in terms of the eigenvalues only.
The dependence of $m_{5q}$ on $N_5$ is discussed in Sec.~\ref{sec:model},
for a given model of the density of the eigenvalues.
In Sec.~\ref{sec:analysis}, using the simplified formula,
we discuss the chiral properties of DWQCD first in the finite volume, and
consider it in the infinite volume limit.
Some comments are also given on the phase structure and the exceptional
configurations in the lattice QCD with the Wilson fermion.
Our conclusion is presented in Sec.~\ref{sec:conclusion}.

%%%%%%%%%%%%%%%%%%%%%%%%%%%%%Section 2%%%%%%%%%%%%%%%%%%%%%%%%%%%%%%%
\section{Action and Axial Ward-Takahashi identity}
\label{sec:action}

We employ Shamir's domain-wall fermion action\cite{Shamir93,Shamir95}. 
Flipping the sign of the Wilson term and the domain wall height $M$, 
we write 
\begin{eqnarray}
S_{DW} &=&
a^4\sum_{x,y,s,s'}
\bpsi(x,s)\biggl[
\left(D_W-\frac{M}{a}\right)_{x,y}\delta_{s,s'}
+\left(D_5\right)_{s,s'}\delta_{x,y}
\biggr]\psi(y,s'),
\\
D_W &=& \sum_\mu\left(
\gamma_\mu\frac{1}{2}\left(\nabla_\mu^f+\nabla_\mu^b\right)
-\frac{a}{2}\nabla_\mu^f\nabla_\mu^b\right),
\label{eqn:Wilson-Dirac}
\\
D_5 &=& \gamma_5\frac{1}{2}\left(\nabla_5^f+\nabla_5^b\right)
-\frac{a_5}{2}\nabla_5^f\nabla_5^b,
\end{eqnarray}
where $x,y$ are four-dimensional space-time coordinates, and $s,s'$ are 
fifth-dimensional or ``flavor'' indices,  bounded as
$1 \le s, s' \le N_5$ with the free boundary condition at both ends
(we assume $N_5$ to be even).
The domain-wall height $M$ should be taken as $0<M<2$ at tree level to
give massless fermion modes.
$\nabla_\mu^f$ and $\nabla_\mu^b$ are forward and backward derivative in 
four dimensions
\begin{eqnarray}
&&
\left(\nabla_\mu^f\right)_{x,y}=
\frac{1}{a}\left(\delta_{x+\mu,y}U_\mu(x)-\delta_{x,y}\right),
\\&&
\left(\nabla_\mu^b\right)_{x,y}=
\frac{1}{a}\left(\delta_{x,y}-\delta_{x-\mu,y}U_\mu^\dagger(y)\right)
\end{eqnarray}
and $\nabla_5^f$ and $\nabla_5^b$ are those in the fifth dimension with
the boundary condition
\begin{eqnarray}
\left(\nabla_5^f\right)_{s,s'}
 &=& \frac{1}{a_5}\left\{
\begin{array}{lll}
\delta_{s+1,s'}  -\delta_{s,s'} &\ & (1\le s<N_5)\\
a_5m_f\delta_{s',1} -\delta_{s,s'} &\ & (s=N_5)
\end{array}
\right.~,
\\
\left(\nabla_5^b\right)_{s,s'}
 &=& \frac{1}{a_5}\left\{
\begin{array}{lll}
\delta_{s,s'}-a_5m_f\delta_{s',N_5} &\ & (s=1)\\
\delta_{s,s'}  -\delta_{s-1,s'}  &\ & (1<s\le N_5)\\
\end{array}
\right.~,
\end{eqnarray}
where $m_f$ is the mass for the quark field.
The light fermion mode is extracted by the 4-dimensional quark field
defined on two boundaries of the fifth dimension,
\begin{eqnarray}
q(x) = P_L \psi(x,1) + P_R \psi(x,{N_5}),
\nn \\
\ovl{q}(x) = \bpsi(x,{N_5}) P_L + \bpsi(x,1) P_R,
\label{eq:quark}
\end{eqnarray}
where $P_{R/L}$ is the projection matrix $P_{R/L}=(1\pm\gamma_5)/2$.
The quark mass term induces a coupling between these boundary fields through 
the bare quark mass $m_f$.

In this paper we investigate chiral property of the domain-wall fermion
through the axial Ward-Takahashi (WT) identity defined in
Ref.~\cite{Shamir95} for the non-singlet axial transformation
\begin{eqnarray}
&&
\delta_A^a \psi(x,s) = i\epsilon(N_5+1-2s) T^a\psi(x,s),
\\&&
\delta_A^a \bpsi(x,s) = -i\epsilon(N_5+1-2s) T^a\bpsi(x,s),
\end{eqnarray}
where $\epsilon (x)$ is a sign function of $x$.
The WT identity for some operator ${\cal O}$ is written as follows. 
\begin{eqnarray}
&&
\vev{\nabla^b_\mu A_\mu^a(x) {\cal O}} =
 2m_f\vev{P^a(x) {\cal O}}
+2\vev{J_{5q}^a(x) {\cal O}}
-\vev{\delta_A^a {\cal O}},
\end{eqnarray}
where an axial vector current $A^a_\mu$, a pseudo scalar 
density $P^a$, and an explicit chiral symmetry breaking term $J_{5q}$
at finite $N_5$ are given by
\begin{eqnarray}
A_\mu^a(x) &=& \sum_{s=1}^{N_5} \epsilon(N_5+1-2s)
\frac{1}{2}\Bigl(
 \bpsi(x,s) T^a(1-\gamma_\mu) U_\mu(x) \psi(x+\mu,s)
\nn\\&&
-\bpsi(x+\mu,s) (1+\gamma_\mu) U_\mu^\dagger(x) T^a \psi(x,s)
\Bigr),
\\
P^a(x) &=& \bq(x)\gamma_5 T^a q(x),
\\
J_{5q}^a(x) &=&
\frac{1}{a_5}\left(
\bpsi(x,\frac{N_5}{2}+1) T^a P_R \psi(x,\frac{N_5}{2})
-\bpsi(x,\frac{N_5}{2}) T^a P_L \psi(x,\frac{N_5}{2}+1)
\right)
\nn\\&=&
\frac{1}{a_5}
\bpsi'(x,\frac{N_5}{2})T^a\gamma_5\psi'(x,\frac{N_5}{2}).
\end{eqnarray}
Here we define the following fields according to \cite{KN99}
\begin{eqnarray}
&&
\psi'(x,\frac{N_5}{2})=P_R\psi(x,\frac{N_5}{2})+P_L\psi(x,\frac{N_5}{2}+1),
\\&&
\bpsi'(x,\frac{N_5}{2})
=\bpsi(x,\frac{N_5}{2}+1)P_R+\bpsi(x,\frac{N_5}{2})P_L.
\end{eqnarray}
In this paper we consider the identity with ${\cal O}=P^b(y)$
\begin{eqnarray}
\vev{\nabla^b_\mu A_\mu^a(x) P^b(y)} &=&
 2m_f\vev{P^a(x) P^b(y)}
+2\vev{J_{5q}^a(x) P^b(y)}
\nn\\&&
-\frac{1}{a^4}\delta_{x,y}\vev{\bq(y)\{T^a,T^b\}q(y)}
\label{eqn:WTidentity}
\end{eqnarray}
and measure the chiral symmetry breaking effect by
\begin{equation}
m_{5q}=\lim_{t\to\infty}
\frac{\sum_{\vec x}\left<J_{5q}(t,{\vec x})P(0)\right>}
        {\sum_{\vec x}\left<P(t,{\vec x})P(0)\right>},
\label{eq:m5q}
\end{equation}
which we call an `anomalous quark mass' \cite{cppacs-dwf}.
Please notice that we have omitted flavor indices since the flavor factors
are canceled in the definition of $m_{5q}$.

%%%%%%%%%%%%%%%%%%%%%%%%%%%%%Section 3%%%%%%%%%%%%%%%%%%%%%%%%%%%%%%%
\section{Effective theory of domain-wall fermions}
\label{sec:effective}

We now derive the effective theory of the DWF system by integrating out the
heavy bulk modes according to Ref.~\cite{KN99}.
Our aim is to rewrite the numerator and denominator of the anomalous 
quark mass \eqn{eq:m5q} in terms of four dimensional quantities and hence
to relate $m_{5q}$ to an hermitian Wilson-Dirac operator in the 
four dimensional theory.
Introducing source fields to $q(x)$, $\bq(x)$,
$\psi'(x,N_5/2)$, $\bpsi'(x,N_5/2)$, 
the propagators necessary for our purpose are derived as\cite{KN99}
\begin{eqnarray}
&&
\vev{q(x)\bq(y)}=
\frac{a_5}{a^5}
\frac{1}{1-am_f}\left(D_{N_5}^{-1}-a\right),
\\&&
\vev{\psi'(x,\frac{N_5}{2})\bq(y)}=
\frac{a_5}{a^5}\frac{1}{2\cosh\frac{N_5}{2}a_5\wt{H}}
D_{N_5}^{-1},
\\&&
\vev{q(x)\bpsi'(y,\frac{N_5}{2})}=
\frac{a_5}{a^5}D_{N_5}^{-1}
\gamma_5\frac{1}{2\cosh\frac{N_5}{2}a_5\wt{H}}\gamma_5,
\end{eqnarray}
%\begin{eqnarray}
%&&
%\vev{q(x)\bq(y)}=
%\frac{a_5}{a^5}
%(1-aD_{N_5})\frac{1}{D_{N_5}+m_f(1-aD_{N_5})},
%\\&&
%\vev{\psi'(x,\frac{N_5}{2})\bq(y)}=
%\frac{a_5}{a^5}
%\frac{1}{2\cosh\frac{N_5}{2}a_5\wt{H}}
%\cdot \frac{1}{D_{N_5}+m_f(1-aD_{N_5})},
%\\&&
%\vev{q(x)\bpsi'(y,\frac{N_5}{2})}=
%\frac{a_5}{a^5}
%\frac{1}{D_{N_5}+m_f(1-aD_{N_5})}
%\gamma_5\frac{1}{2\cosh\frac{N_5}{2}a_5\wt{H}}\gamma_5,
%\end{eqnarray}
where $\wt{H}$ is a DWF Hamiltonian in the 5th direction, 
with which the transfer matrix in fifth direction is given by
\begin{eqnarray}
 T=e^{-a_5\wt{H}}.
\end{eqnarray}
$D_{N_5}$ is a truncated Ginsparg-Wilson (GW) Dirac operator which
satisfies GW relation in $N_5\to\infty$ limit.
$\wt{H}$ and $D_{N_5}$ are related to the four dimensional hermitian
Wilson-Dirac operator $H_W$ as,
\begin{eqnarray}
&&
D_{N_5} = \frac{1}{2a}\left[(1+a m_f)+(1-a m_f)
\gamma_5\tanh\frac{N_5}{2}a_5\wt{H}\right],
\label{eqn:DN5}
\\&&
\wt{H} =\frac{1}{a_5}\log \frac{1+aH'}{1-aH'},
\\&&
H'=H_W\frac{1}{2+a\gamma_5H_W},
\label{eqn:H'}
\\&&
H_W=\gamma_5\left(D_W-\frac{M}{a}\right).
\end{eqnarray}
Here we adopted Bori\c{c}i's notation\cite{Borici} for $\wt{H}$.

The numerator of $m_{5q}$ can be written in
terms of the above propagators,
\begin{eqnarray}
X(t) &=& \sum_{\vec{x}}\vev{J_{5q}(\vec{x},t)P(\vec{y},0)}
\nn\\&=&
-\frac{1}{a_5}\sum_{\vec{x}}
\tr\left[\gamma_5\vev{\psi'(\vec{x},t,\frac{N_5}{2})\bq(\vec{y},0)}\gamma_5
 \vev{q(\vec{y},0)\bpsi'(\vec{x},t,\frac{N_5}{2})}
\right]
\nn\\&=&
-\frac{a_5}{a^{10}}\sum_{\vec{x},\alpha,a}\sum_{\beta,b}
\VEV{I}{\frac{1}{2\cosh\frac{N_5}{2}a_5\wt{H}}D_{N_5}^{-1}}{J}
\VEV{J}{\left(D_{N_5}^{-1}\right)^\dagger
\frac{1}{2\cosh\frac{N_5}{2}a_5\wt{H}}}{I},
\label{eqn:numerator}
\end{eqnarray}
where $I=(\vec{x},t,\alpha,a)$, $J=(\vec{y},0,\beta,b)$. $\alpha,\beta$ are
spinor indices and $a,b$ are color indices.
$\ket{I}$ is an eigen-ket in the coordinate, spinor and color spaces.
In the last equality we use a relation that
$\gamma_5(D_{N_5}^{-1})\gamma_5=(D_{N_5}^{-1})^\dagger$.
The denominator of $m_{5q}$ is given by
\begin{eqnarray}
Y(t) &=& \sum_{\vec{x}}\vev{P(\vec{x},t)P(\vec{y},0)}
\nn\\&=&
-\sum_{\vec{x}}\tr\left[\gamma_5\vev{q(\vec{x},t)\bq(\vec{y},0)}
\gamma_5\vev{q(\vec{y},0)\bq(\vec{x},t)}\right]
\nn\\&=&
-\frac{a_5^2}{a^{10}}\frac{1}{(1-a m_f)^2}\sum_{\vec{x},\alpha,a}\sum_{\beta,b}
\VEV{I}{\left(D_{N_5}^{-1}-a\right)}{J}
\VEV{J}{\gamma_5\left(D_{N_5}^{-1}-a\right)\gamma_5}{I}
\nn\\&=&
-\frac{a_5^2}{a^{10}}\frac{1}{(1-a m_f)^2}\sum_{\vec{x},\alpha,a}\sum_{\beta,b}
\VEV{I}{D_{N_5}^{-1}}{J}
\VEV{J}{\left(D_{N_5}^{-1}\right)^\dagger}{I}.
\label{eqn:denominator}
\end{eqnarray}
Note that $\braket{I}{J}=0$ for $t\neq0$.

Hereafter we set the bare quark mass $m_f=0$ without loss of
generality.

%%%%%%%%%%%%%%%%%%%%%%%%%%%%%Section 7%%%%%%%%%%%%%%%%%%%%%%%%%%%%%%%
\section{Expansion in terms of eigenstates}
\label{sec:expansion}

We expand $X(t)$ and $Y(t)$ in terms of eigenstates of the
hermitian operator $H'$.
Since the numerator \eqn{eqn:numerator} has a suppression factor
$1/(2\cosh\frac{N_5}{2}a_5\wt{H})$ only small eigenvalues of $\wt{H}$,
or equivalently those of $H'$, contribute to $m_{5q}$ at large $N_5$.
In this case we can expand $H'$ perturbatively in terms of the four
dimensional hermitian Wilson-Dirac operator $H_W$ and can derive the
formula which connect $m_{5q}$ and the eigenvalue of $H_W$ at large
$N_5$.

The expansion of $X(t)$ and $Y(t)$ with these eigenstates is done by
inserting a complete set of eigenstates
\begin{eqnarray}
1=\sum_{n}\ket{n}\bra{n},
\end{eqnarray}
where $\ket{n}$ is $n$-th eigenstate of $H'$
\begin{eqnarray}
H'\ket{n}=\lambda'_n\ket{n}.
\end{eqnarray}
With this substitution $X(t)$ and $Y(t)$ are written in terms of
eigenvalue and eigenfunction.

In order to further simplify these Green functions as a function of 
eigenvalue only, we employ the following assumptions and
approximations for typical form of the eigenfunctions.
The recent numerical analyses indicate that
the eigenvalues of $H_W$ seem to be classified into two groups
\cite{HJL98,Nagai00}; one is a group of small isolated eigenvalues
and the other is a group of almost continuous eigenvalues above the
isolated ones.
The eigenvectors associated with the isolated eigenvalues are exponentially
localized at some center in the coordinate space \cite{HJL98,Nagai00}, and
those for the continuous eigenvalues are rapidly oscillating plane-wave
functions in the coordinate space\cite{Nagai00}.
This property of eigenvalues is expected to hold also for $H'$,
since $H'$ can be expanded in terms of $H_W$ perturbatively for small 
eigenvalues.
>From this consideration
we assume that the eigenvector space of $H'$ is divided
into two subspaces ${\cal S}$ and $\wt{\cal S}$, where
${\cal S}$ is spanned by localized eigenvectors and $\wt{\cal S}$ is
expanded with plane wave functions.

\subsection{Localized eigenvectors}

We adopt two types of the approximation for typical form of the
localized eigenvectors; 
completely localized ones and partially localized ones.
The completely localized eigenvector means that the eigenfunction
$\psi_n$ of the $n$-th eigenvalue has non-zero value at a single parameter
$(x_n,\alpha_n,a_n)$ 
\begin{eqnarray}
 \psi_n(x,\alpha,a)=\delta_{x,x_n}\delta_{\alpha,\alpha_n}\delta_{a,a_n},
\end{eqnarray}
which gives a unit vector in $(x,\alpha,a)$ space.
$\psi_n(x,\alpha,a)$ is given by using the $n$-th eigenstate $\ket{n}$
of an eigenvalue $\lambda'_n$
\begin{eqnarray}
&&
\psi_n(x,\alpha,a)=\braket{x,\alpha,a}{n},
\\&&
H'\ket{n}=\lambda_n'\ket{n},
\\&&
\wt{H}\ket{n}=\wt{\lambda}(\lambda_n')\ket{n},
\\&&
\wt{\lambda}(\lambda_n') =\frac{1}{a_5}\log
 \frac{1+a\lambda'_n}{1-a\lambda_n'}.
\end{eqnarray}
We assume that eigenvectors with different eigenvalues reside 
at different points
\begin{eqnarray}
 \braket{n}{x,\alpha,a}\braket{x,\alpha,a}{m}=0\quad
{\rm for}\; n \neq m.
\end{eqnarray}
In this case ${\cal S}$ is spanned by a set of basis vectors
\begin{eqnarray}
 \pmatrix{1\cr0\cr\vdots},\quad
 \pmatrix{0\cr1\cr\vdots},\quad\cdots.
\end{eqnarray}

On the other hand, the partially localized eigenvector is non-zero at
small range of $(x,\alpha,a)$ space.
We call this small range of volume $h_n$ as ${\cal S}_n$ and
${\cal S}=\cup_n{\cal S}_n$.
${\cal S}_n$ is spanned by a set of eigenvectors $\psi_n^{(i)}$,
where $i$ runs as $i=1,\cdots,h_n$ and $\psi_n^{(i)}$ is given by
\begin{eqnarray}
&&
\psi_n^{(i)}(x,\alpha,a)=\braket{x,\alpha,a}{n,i},
\\&&
H'\ket{n,i}=\lambda_n^{'(i)}\ket{n,i}.
\end{eqnarray}
We assume that ${\cal S}_n$ with different $n$ has no overlap each
other; 
\begin{eqnarray}
\braket{m,i}{x,\alpha,a}\braket{x,\alpha,a}{n,j}=0
\end{eqnarray}
for $m \neq n$ for any $i,j$.
In the limit of $h_n\rightarrow 1$ the partially localized eigenstate
is reduced to the completely localized one.

Correspondence between the eigenvalues and eigenvectors of $H'$ and
$H_W$ is given perturbatively for small eigenvalues.
We expand $H'$ in terms of $H_W$
\begin{eqnarray}
&&
2 H' = H_W+H_W\sum_{k=1}^{\infty}(-1)^{k} 
\left(\frac{a\gamma_5H_W}{2}\right)^{k},
\end{eqnarray}
and the eigenvalue $\lambda_n^W$ and eigenstate $\ket{n_W}$ of $H_W$ is
defined by
\begin{eqnarray}
H_W\ket{n_W}=\lambda^W_n\ket{n_W}.
\end{eqnarray}
The standard perturbation theory gives
the relation between $\lambda_n$, $\ket{n}$ and $\lambda_n^W$,
$\ket{n_W}$ as follows.
\begin{eqnarray}
&&
2\lambda'_n=\lambda_n^W
-\frac{1}{2}\left(\lambda_n^W\right)^2\VEV{n_W}{\gamma_5}{n_W}
+{\cal O}(\lambda^3),
\\&&
\ket{n}=\ket{n_W}
-\frac{1}{2}\lambda_n^W\frac{\phi_n}{\lambda_n^W-H_W}H_W \gamma_5 \ket{n_W}
+{\cal O}(\lambda^3),
\end{eqnarray}
where $\phi_n$ is a projection operator onto the space perpendicular
to the $n$-th eigenstate:
\begin{eqnarray}
\phi_n=1-\ket{n_W}\bra{n_W}.
\end{eqnarray}

\subsection{Plane-wave eigenvector}

We approximate the plane-wave function with that of the free theory
but relax the condition for possible momenta $p$ in the finite box.
The operator $H'$ of the free theory in momentum space is given by
\begin{eqnarray}
&&
H'(p)=\frac{1}{a}\gamma_5\left(i\sum_\mu\gamma_\mu A_\mu(p)+B(p)\right),
\label{eqn:free-H}
\\&&
A_\mu(p) = \frac{2s_\mu}{s^2+(2+2\hat{s}^2-M)^2},
\\&&
B(p)=\frac{s^2+(2\hat{s}^2-M)(2+2\hat{s}^2-M)}{s^2+(2+2\hat{s}^2-M)^2},
\end{eqnarray}
where $s_\mu=\sin (a p_\mu )$,
$\hat{s}^2=\sum_\mu\sin^2(ap_\mu/2)$ and $s^2=\sum_\mu s_\mu^2$.
The eigenstate of free $H'$ and $\wt{H}$ becomes
\begin{eqnarray}
&&
H'\ket{p,s,a}=\lambda'(p)\ket{p,s,a},
\\&&
\wt{H}\ket{p,s,a}=\wt{\lambda}(p)\ket{p,s,a},
\\&&
\lambda'(p)=\pm\frac{1}{a}\sqrt{A^2(p)+B^2(p)},
\\&&
\wt{\lambda}(p)=
\frac{1}{a_5}\log\frac{1+a\lambda'(p)}{1-a\lambda'(p)},
\end{eqnarray}
where the spinor index $s$ runs as $s=1,2,3,4$, and the eigenstate is
degenerate in the color index $a$.
The eigenvector in $(x,\alpha,a)$ space can be written as
\begin{eqnarray}
\braket{x,\alpha,a}{p,s,b}=
\frac{1}{\sqrt{N_p}}U_\alpha(p,s)e^{ipx}\delta_{a,b}
\label{eqn:plane-wave}
\end{eqnarray}
within the subspace $\wt{\cal S}$, where $U_\alpha(p,s)$ is a normalized
eigenfunction of free $H'(p)$, given in Appendix \ref{sec:appendix-a},
and $N_p$ is the total number of the plane-wave eigenvectors.
The free truncated overlap Dirac operator and its inverse in momentum space are
given by
\begin{eqnarray}
\VEV{k,s,a}{D_{N_5}}{p,t,b} &=&
\frac{1}{2a}
\left(i(\gamma_\mu)_{s,t}C_\mu(p) +\delta_{s,t}E(p)\right)
 \delta_{k,p}\delta_{a,b},
\\
\VEV{k,s,a}{D_{N_5}^{-1}}{p,t,b} &=&
2a\frac{-i(\gamma_\mu)_{s,t}C_\mu(p)+\delta_{s,t}E(p)}{C^2+E^2}
 \delta_{k,p}\delta_{a,b},
\end{eqnarray}
where
\begin{eqnarray}
&&
C_\mu(p)=\frac{A_\mu(p)}{a|\lambda'(p)|}\tanh\frac{N_5}{2}a_5|\wt{\lambda}(p)|,
\\&&
E(p)=\left(1+\frac{B(p)}{a|\lambda'(p)|}
 \tanh\frac{N_5}{2}a_5|\wt{\lambda}(p)|\right).
\end{eqnarray}

In addition 
we also assume that the overlap of two eigenfunctions from different groups
vanishes: 
\begin{eqnarray}
\braket{n}{x,\alpha,a}\braket{x,\alpha,a}{p,s,b}=0.
\end{eqnarray}

The eigenvalues of the hermitian Wilson-Dirac operator $H_W$ is also given as a
function of $p_\mu$ for the free theory.
The free $H_W$ in the momentum space is given by
\begin{eqnarray}
H_W&=&
\gamma_5\left(i\gamma_\mu s_\mu+2\hat{s}^2-M\right).
\end{eqnarray}
The eigenvalue $\lambda_W(p)$ of free $H_W$ is given by
\begin{eqnarray}
&&
H_W\ket{\lambda_W(p)}=\lambda_W(p)\ket{\lambda_W(p)},
\\&&
\lambda_W(p)=\pm\sqrt{s^2+\left(2\hat{s}^2-M\right)^2},
\end{eqnarray}
where $\ket{\lambda_W(p)}$ is the corresponding eigenstate.
Although the eigenstates of $\lambda'(p)$ and $\lambda_W(p)$
are different, $\lambda'(p)$ of $H'$ is given in terms of $\lambda_W(p)$ 
and the momentum $p$ as
\begin{eqnarray}
\lambda'(p)&=&
\frac{\lambda_W(p)}{\sqrt{\lambda_W(p)^2+4(2\hat{s}^2-M)+4}}.
\end{eqnarray}

\subsection{Simplification of the formula of $m_{5q}$}

We first consider a system with $N_l$ completely localized eigenvectors and
plane wave functions with $N_p$ degrees of freedom for the momentum $p$.
$N_l$ and $N_p$ satisfy a relation $N_l+4N_cN_p=4N_cN_x$, where $N_x$ is
a number of total sites of our lattice $N_x=n_xn_yn_zn_t$.
Since the total eigenvector space is separated into two subspaces
${\cal S}$ and $\wt{\cal S}$, the complete set of this system is given
by
\begin{eqnarray}
1=\sum_{n=1}^{N_l}\ket{n}\bra{n}
+\sum_{p=1}^{N_p}\sum_{s=1}^4\sum_{a=1}^{N_c}\ket{p,s,a}\bra{p,s,a},
\label{eqn:complete-1}
\end{eqnarray}
where $\ket{n}\in{\cal S}$ and $\ket{p,s,a}\in\wt{\cal S}$.

We expand the numerator of $m_{5q}$ with
the complete set in the above.
\begin{eqnarray}
X(t) &=&
% \sum_{\vec{x}}\vev{J_{5q}(x)P(y)}
-\frac{a_5}{a^{10}}\sum_{\vec{x},\alpha,a}\sum_{b,\beta}
\VEV{I}{\frac{1}{2\cosh\frac{N_5}{2}a_5\wt{H}}D_{N_5}^{-1}}{J}
\VEV{J}{\left(D_{N_5}^{-1}\right)^\dagger
\frac{1}{2\cosh\frac{N_5}{2}a_5\wt{H}}}{I}
\nn\\&=&
X_l(t) + X_c(t),
\end{eqnarray}
where $X_l(t)$ and $X_c(t)$ are the contributions from 
the localized eigenvectors and the plane wave function, respectively.
After a little algebra, the detail of which is given in 
Appendix \ref{sec:appendix-b},
we obtain 
\begin{eqnarray}
X_l(t) &=&
\frac{1}{a_5}Y(t)\frac{1}{4N_cN_x}\sum_{n=1}^{N_l}
\left(\frac{1}{2\cosh\frac{N_5}{2}a_5\wt{\lambda}(\lambda'_n)}\right)^2
\end{eqnarray}
for the contribution from the localized eigenvectors,
where $Y(t)$ is the denominator of $m_{5q}$.

Similarly, taking $t\to\infty$ limit, we have for the contribution of the  
continuous eigenvalues
\begin{eqnarray}
X_c(t)|_{t\to\infty} &=&
-\frac{a_5}{a^{8}}n_{xyz}'\frac{1}{N_p}\sum_{p}
\left(\frac{1}{2\cosh\frac{N_5}{2}a_5\wt{\lambda}(p)}\right)^2
F(p),
\end{eqnarray}
where $n'_{xyz}$ is a volume of $\wt{\cal S}$ in $xyz$ space, and
$F(p)$ is defined as
\begin{eqnarray}
F(p)=
%\sum_{n=1}^{N_l}\delta_{y,x_n}\sum_{s,a}
%\frac{1}{a^2}
%\VEV{p,s,a}{D_{N_5}^{-1}}{n}\VEV{n}{\left(D_{N_5}^{-1}\right)^\dagger}{p,s,a}
\frac{4N_c}{N_p}\left(\frac{4}{C^2(p)+E^2(p)}\right).
\end{eqnarray}

We next evaluate the denominator of $m_{5q}$ in the similar manner:
\begin{eqnarray}
Y(t) &=&
%\sum_{\vec{x}}\vev{P(x)P(y)}
-\frac{a_5^2}{a^{10}}\sum_{\vec{x},\alpha,a}\sum_{\beta,b}
\VEV{I}{D_{N_5}^{-1}}{J}
\VEV{J}{\left(D_{N_5}^{-1}\right)^\dagger}{I}
=
 Y_l(t) + Y_c(t),
\end{eqnarray}
where
\begin{eqnarray}
Y_l(t)&=&\frac{N_l}{4N_cN_x}Y(t)
\end{eqnarray}
and
\begin{eqnarray}
Y_c(t)|_{t\to\infty} &=&
-\frac{a_5^2}{a^{10}}n_{xyz}'\frac{1}{N_p}
\sum_{p}F(p).
\end{eqnarray}
These results imply
\begin{eqnarray}
Y(t\to\infty)= -\frac{4N_cN_x}{4N_cN_x-N_l}
\frac{a_5^2}{a^{10}}n_{xyz}'
\frac{1}{N_p}\sum_{p}F(p).
\end{eqnarray}

\begin{figure}[htb]
 \begin{center}
  \leavevmode
  \epsfxsize=7cm \epsfbox{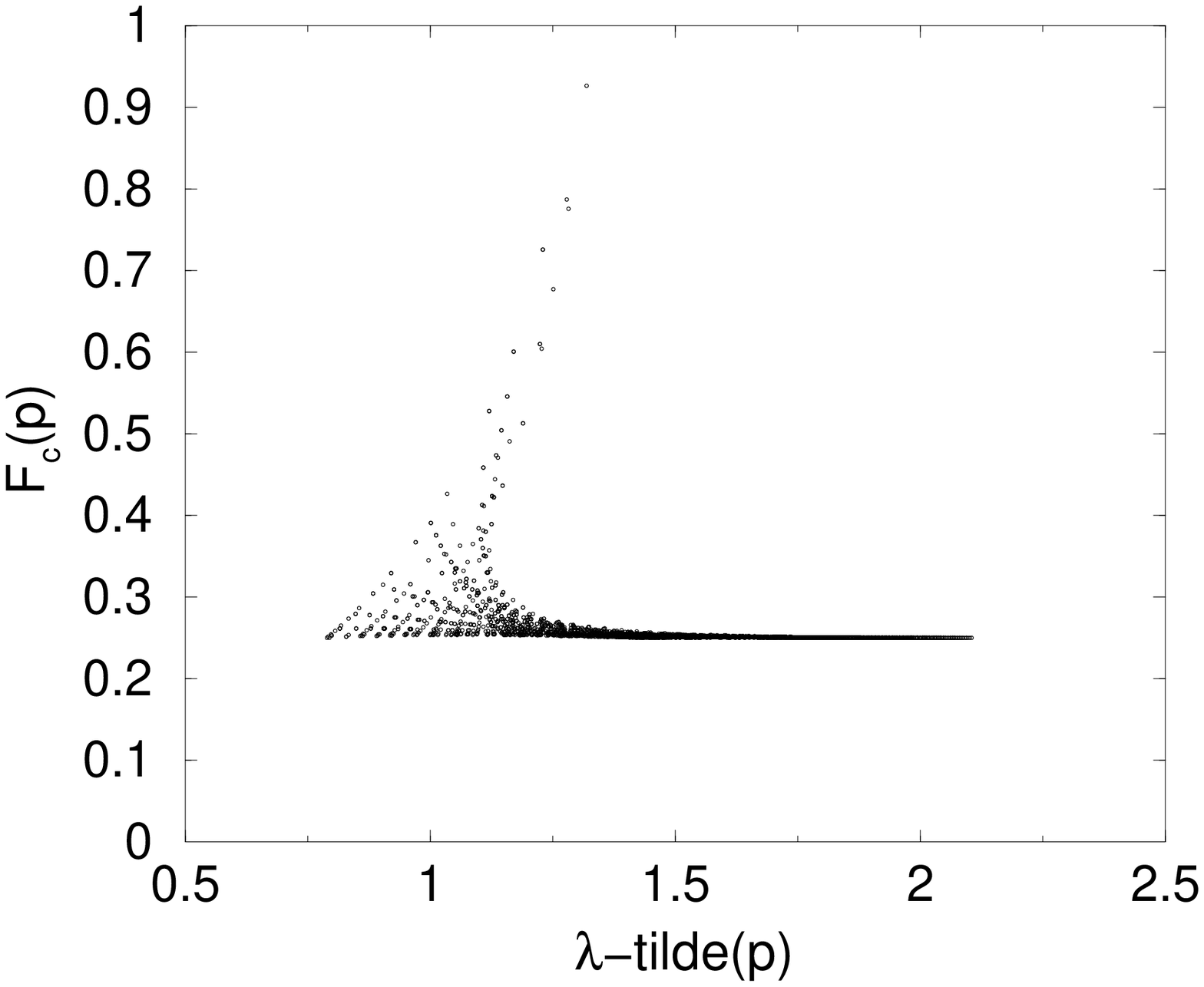}
  \epsfxsize=7cm \epsfbox{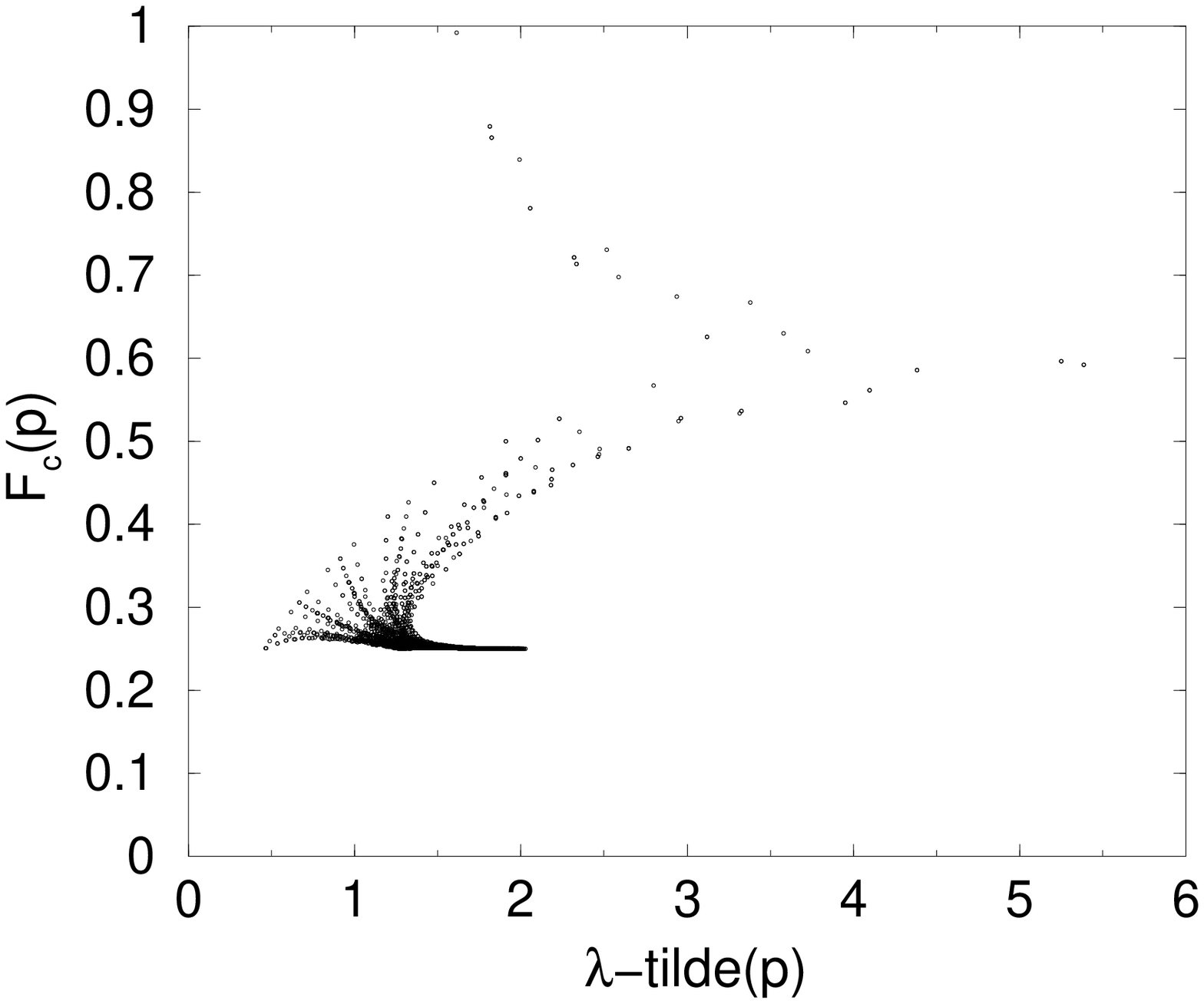}
\caption{$F_c(p)$ as a function of $\wt{\lambda}(p)$
at $M=0.8$(left) and $M=1.4$(right). Both $F_c$ and
  $\wt{\lambda}$ are evaluated on $16^3\times32$ momentum lattice,
where the periodic boundary condition is adopted for
  spatial direction and anti-periodic boundary condition is set for
  temporal direction in order to prevent a singularity at the origin.}
\label{fig:Fp}
 \end{center}
\end{figure}

%The explicit form of $F(p)$ as a function of $p$
%is given by
%\begin{eqnarray}
%F(p) &=& F_l(p)+F_c(p),
%\\
%F_l(p) &=& \sum_{n=1}^{N_l}\delta_{y,x_n}\sum_{s,a}\frac{1}{a^2}
%\VEV{p,s,a}{D_{N_5}^{-1}}{n}\VEV{n}{\left(D_{N_5}^{-1}\right)^\dagger}{p,s,a},
%\\
%F_c(p) &=& \frac{4N_c}{N_p}\left(\frac{4}{C^2(p)+E^2(p)}\right).
%\end{eqnarray}
We can directly see that the function $F(p)$ is almost constant for
all the range of $p$ as in Fig.~\ref{fig:Fp} except some discrete points 
of $p$. Therefore it is a reasonable approximation to
assume that $F(p)$ is independent of $p$.
Combining these formula and approximations
the anomalous quark mass becomes
\begin{eqnarray}
m_{5q}&=&\frac{\lim_{t\to\infty}X(t)}{\lim_{t\to\infty}Y(t)}
\nn\\&=&
\frac{1}{a_5}\frac{1}{4N_cN_x}
\left(
\sum_{n}
\left(\frac{1}{2\cosh\frac{N_5}{2}a_5\wt{\lambda}(\lambda'_n)}\right)^2
+\sum_{n}\rho_n\left(\frac{1}{2\cosh\frac{N_5}{2}a_5\wt{\lambda}_n}\right)^2
\right),
\end{eqnarray}
where $\rho_n$ is a number of degeneracy in $\wt{\lambda}_n$,
\begin{eqnarray}
\rho_n=\sum_{p,\alpha,a}\delta_{\wt{\lambda}_n,\wt{\lambda}(p)}.
\end{eqnarray}
If the degeneracy of $\rho_n$ in the free case is resolved by
the presence of gauge fields,
we are able to reconstruct $m_{5q}$ by simply summing up all eigenvalues
of $H'$,
\begin{eqnarray}
 m_{5q}=\frac{1}{a_5}\frac{1}{4N_cN_x}\sum_{n}
\left(\frac{1}{2\cosh\frac{N_5}{2}a_5\wt{\lambda}(\lambda'_n)}\right)^2.
\end{eqnarray}

\vskip 1.0cm

The above result can be generalized to the case with the partially
localized eigenstates. The formula becomes
\begin{eqnarray}
m_{5q}&=&\frac{\lim_{t\to\infty}X(t)}{\lim_{t\to\infty}Y(t)}
\nn\\&=&
\frac{1}{a_5}\frac{1}{4N_cN_x}
\left(
\sum_{n=1}^{N_l}\sum_{i=1}^{h_n}\tilde h_n^{(i)}
\left(\frac{1}{2\cosh\frac{N_5}{2}a_5\wt{\lambda}(\lambda^{'(i)}_n)}\right)^2
+
\sum_{n}\rho_n\left(\frac{1}{2\cosh\frac{N_5}{2}a_5\wt{\lambda}_n}
\right)^2\right),
\end{eqnarray}
where the $n$-th set of localized eigenvectors have the (local) support
with the dimensionless volume of $h_n$, and the $\tilde h_n^{(i)}$
 ($1\le \tilde h_n^{(i)}\le 4^4h_n$)
is the enhancement factor for the set, which depends on the shape of
eigenvectors.
See the detail of the derivation in  Appendix \ref{sec:appendix-c}.
When the degeneracy in the continuous eigenvalues is resolved with gauge
fields we have
\begin{eqnarray}
&&
m_{5q}a_5=
\frac{1}{4N_cN_x}\left(
\sum_{\rm local} \tilde h_n
\left(\frac{1}{2\cosh\frac{N_5}{2}a_5\wt{\lambda}(\lambda'_n)}\right)^2
+\sum_{\rm continuous}
\left(\frac{1}{2\cosh\frac{N_5}{2}a_5\wt{\lambda}(\lambda'_n)}\right)^2
\right).
\nn\\
\end{eqnarray}

%%%%%%%%%%%%%%%%%%%%%%%%%%%%%Section 7%%%%%%%%%%%%%%%%%%%%%%%%%%%%%%%
\section{Model for eigenvalue density}
\label{sec:model}

\begin{figure}[htb]
 \begin{center}
  \leavevmode
  \epsfxsize=5cm \epsfbox{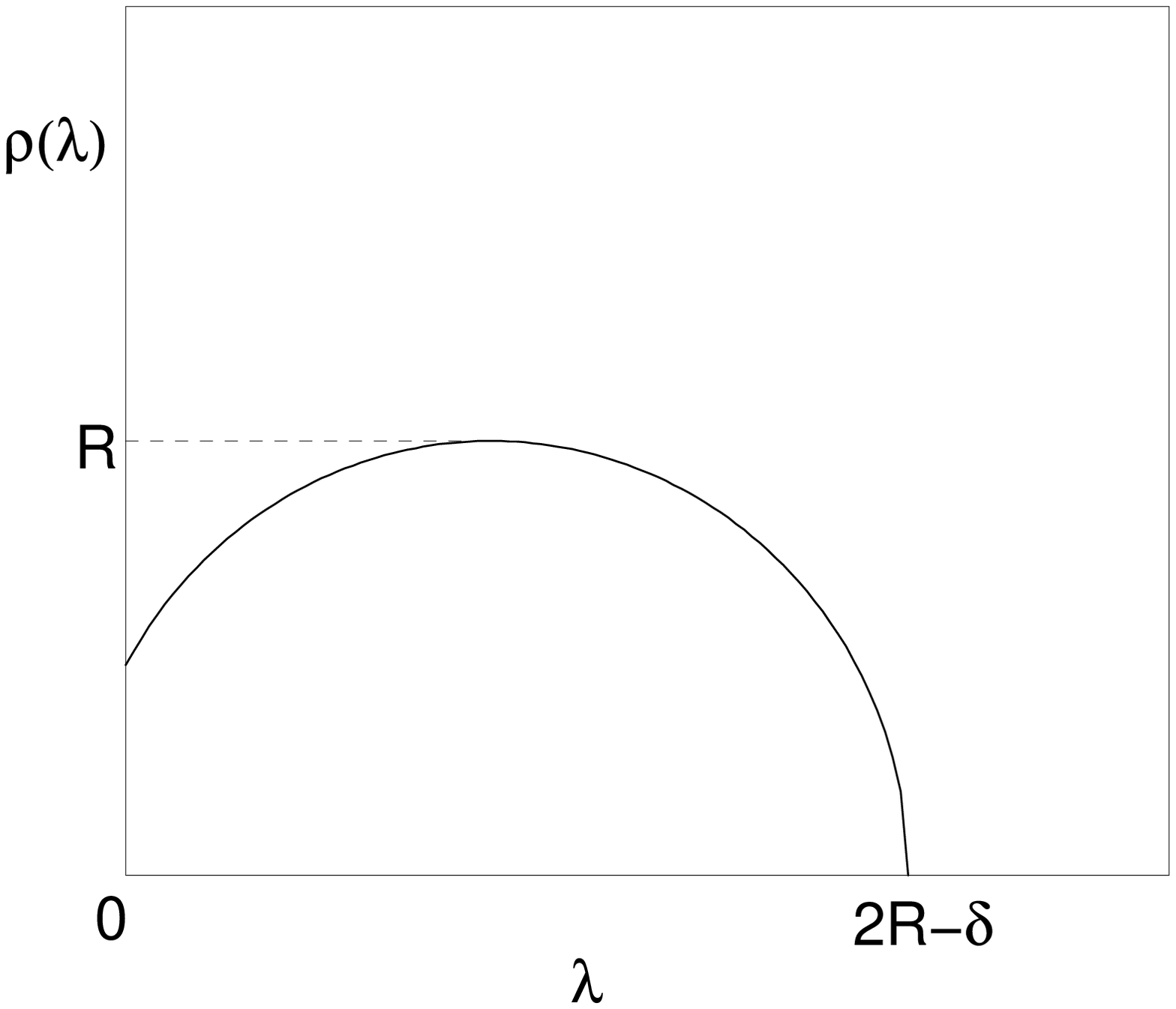}
  \epsfxsize=5cm \epsfbox{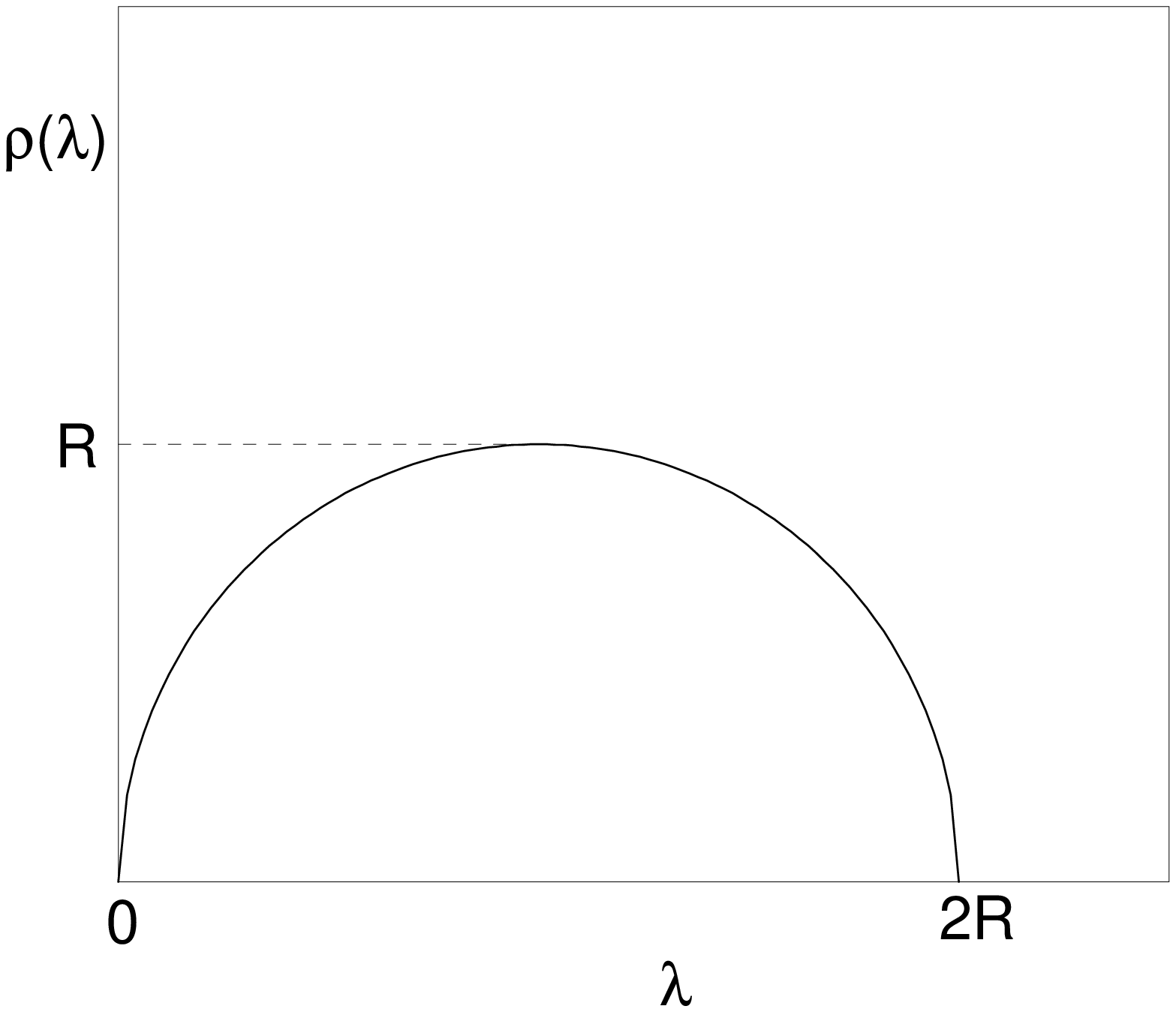}
  \epsfxsize=5cm \epsfbox{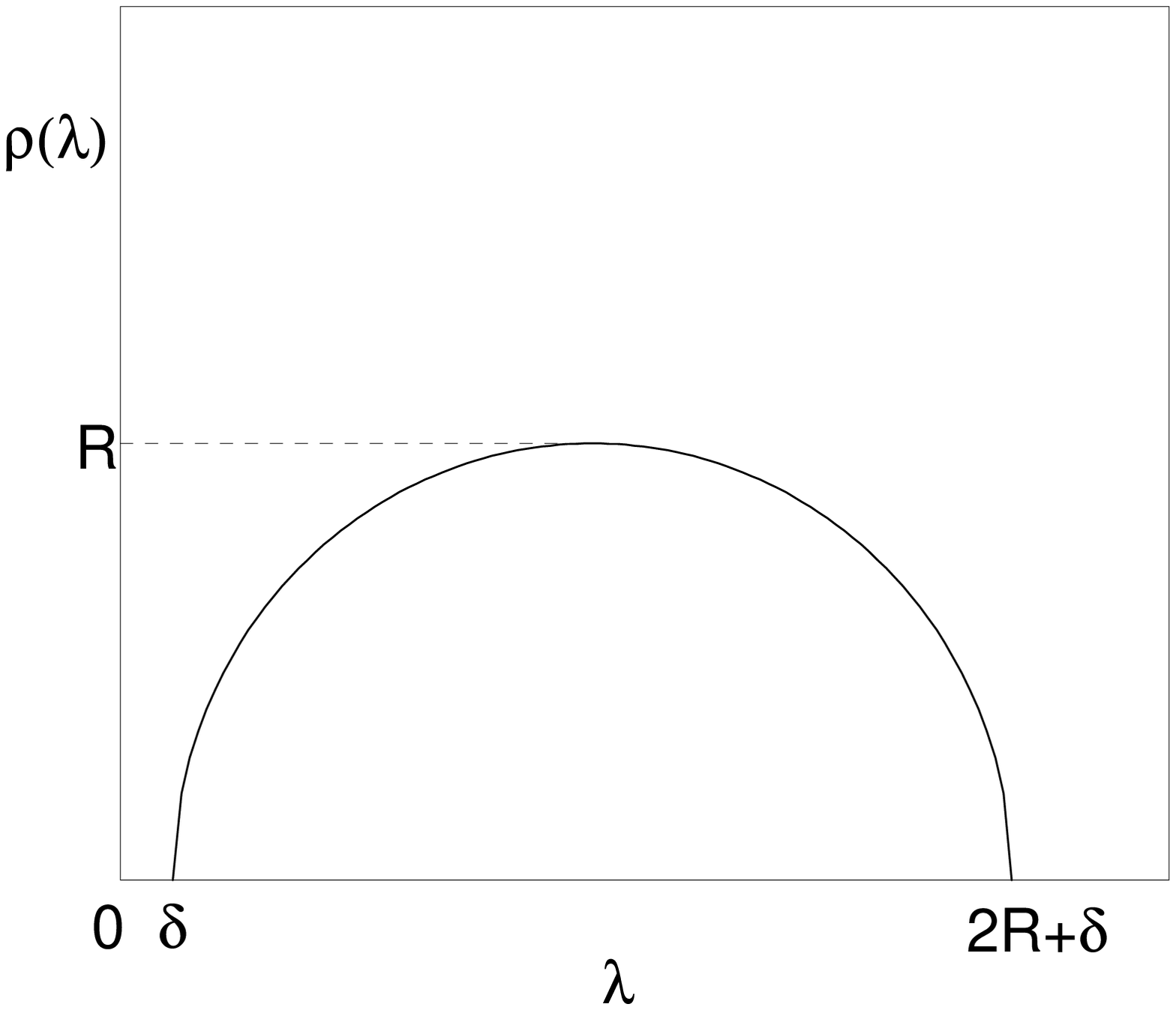}
\caption{Three types of $\rho(\lambda)$ adopted in calculation of
  $m_{5q}$. The left one corresponds to type (1) with $\delta<0$,
  the center is type (2) with $\delta=0$ and the right is type (3) with
  $\delta>0$.} 
\label{fig:rho}
 \end{center}
\end{figure}

In this section
we consider the relation between the asymptotic behavior of $m_{5q}$ in 
$N_5$ and the distribution of the continuous eigenvalues. For simplicity 
we neglect the effect
of the localized eigenvalues here and will discuss their effect in the
next section.
In this simple situation
we can write $m_{5q}$ as an integral in continuous eigenvalues
\begin{eqnarray}
m_{5q}a_5=
\int_{\lambda_{\rm min}}^{\lambda_{\rm max}}d\lambda
\rho(\lambda)
\left(\frac{1}{2\cosh\frac{N_5}{2}\lambda}\right)^2,
\end{eqnarray}
where $\lambda$ is a dimensionless eigenvalue
$\lambda=a_5\wt{\lambda}$,
$\lambda_{\rm min}$ and $\lambda_{\rm max}$ are minimum and maximum of
eigenvalues $0\le\lambda_{\rm min}\le\lambda\le\lambda_{\rm max}$.
Without loss of generality, we consider non-negative $\lambda$ only, by taking 
$\rho(\lambda)=\rho_n(\lambda)+\rho_n(-\lambda)$,
since $\cosh\frac{N_5}{2}\lambda$ is the even function.
We adopt the following three types of $\rho(\lambda)$
\begin{eqnarray}
\rho(\lambda)=\sqrt{R^2-(\lambda-R-\delta)^2}
\end{eqnarray}
with
\begin{enumerate}
 \item $\delta<0$, $\lambda_{\rm min}=0$, $\lambda_{\rm max}=2R-|\delta|$,
       $\rho(0)\neq0$,
 \item $\delta=0$, $\lambda_{\rm min}=0$, $\lambda_{\rm max}=2R$,
       $\rho(0)=0$,
 \item $\delta>0$, $\lambda_{\rm min}=\delta$, $\lambda_{\rm max}=2R+\delta$,
       $\rho(0)=0$.
\end{enumerate}
A typical form of $\rho(\lambda)$ for each case is given in Fig.~\ref{fig:rho}

A main support of the integral at large $N_5$ resides near
$\lambda_{\rm min}$ and asymptotic behavior of $m_{5q}$ is evaluated by
expanding $\rho(\lambda)$ around $\lambda_{\rm min}$ and adopting the
leading term.
Since contribution from the larger eigenvalues is negligible we set the
integral range to $[\lambda_{\rm min},\infty]$.
The integral in $\lambda$ is easily calculated with the following two
formulas,
\begin{eqnarray}
&&
\left(\frac{1}{2\cosh\frac{N_5}{2}\lambda}\right)^2=
\sum_{n=1}^{\infty}(-)^{n-1}ne^{-nN_5\lambda},
\\&&
\int_{0}^{\infty}d\lambda e^{-x\lambda}\lambda^\alpha
=\frac{\Gamma(\alpha+1)}{x^{\alpha+1}},
\end{eqnarray}
which is valid for $x>0$, $\alpha>-1$.

\begin{figure}[htb]
 \begin{center}
  \leavevmode
  \epsfxsize=5.5cm \epsfbox{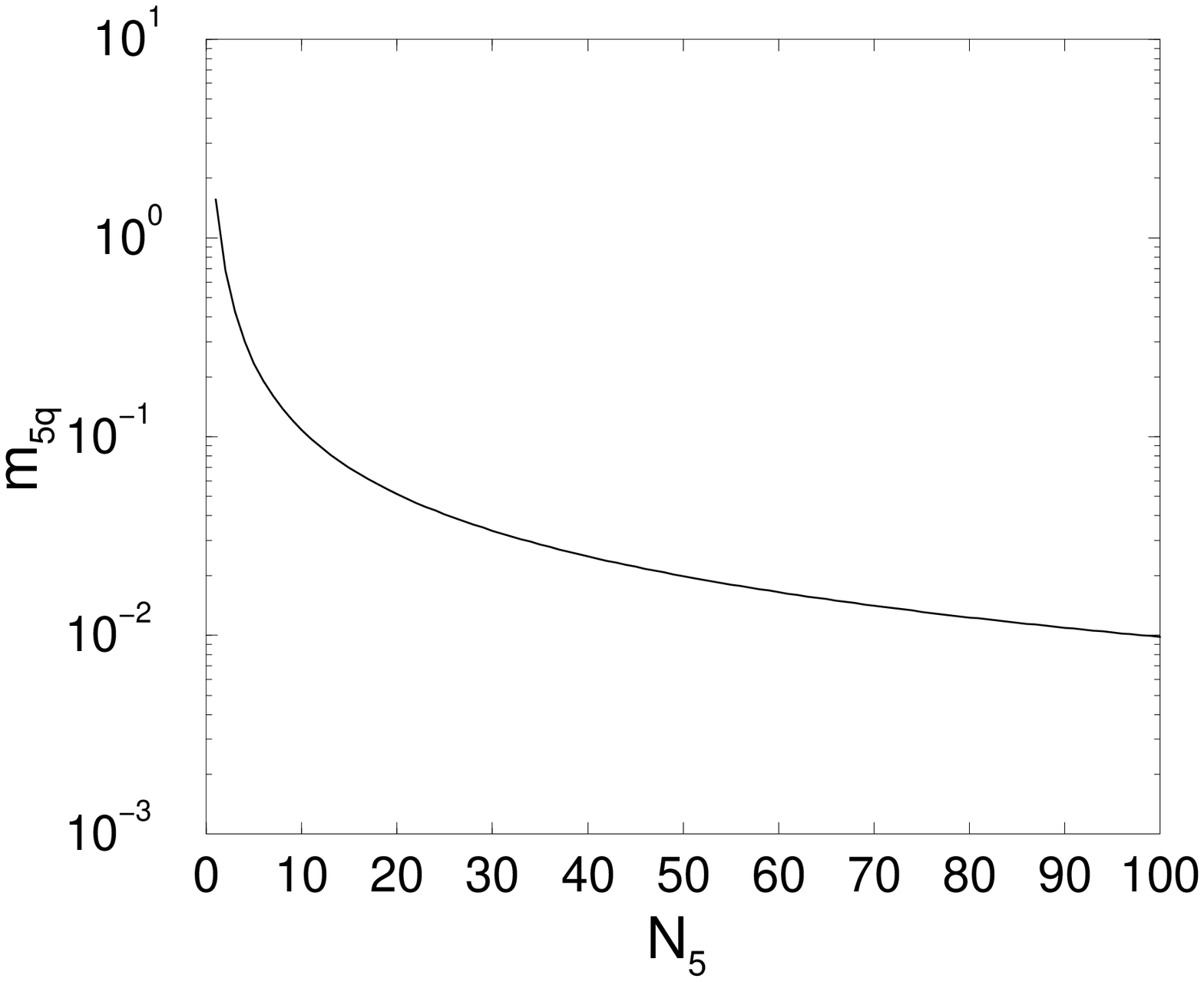}
  \epsfxsize=5.5cm \epsfbox{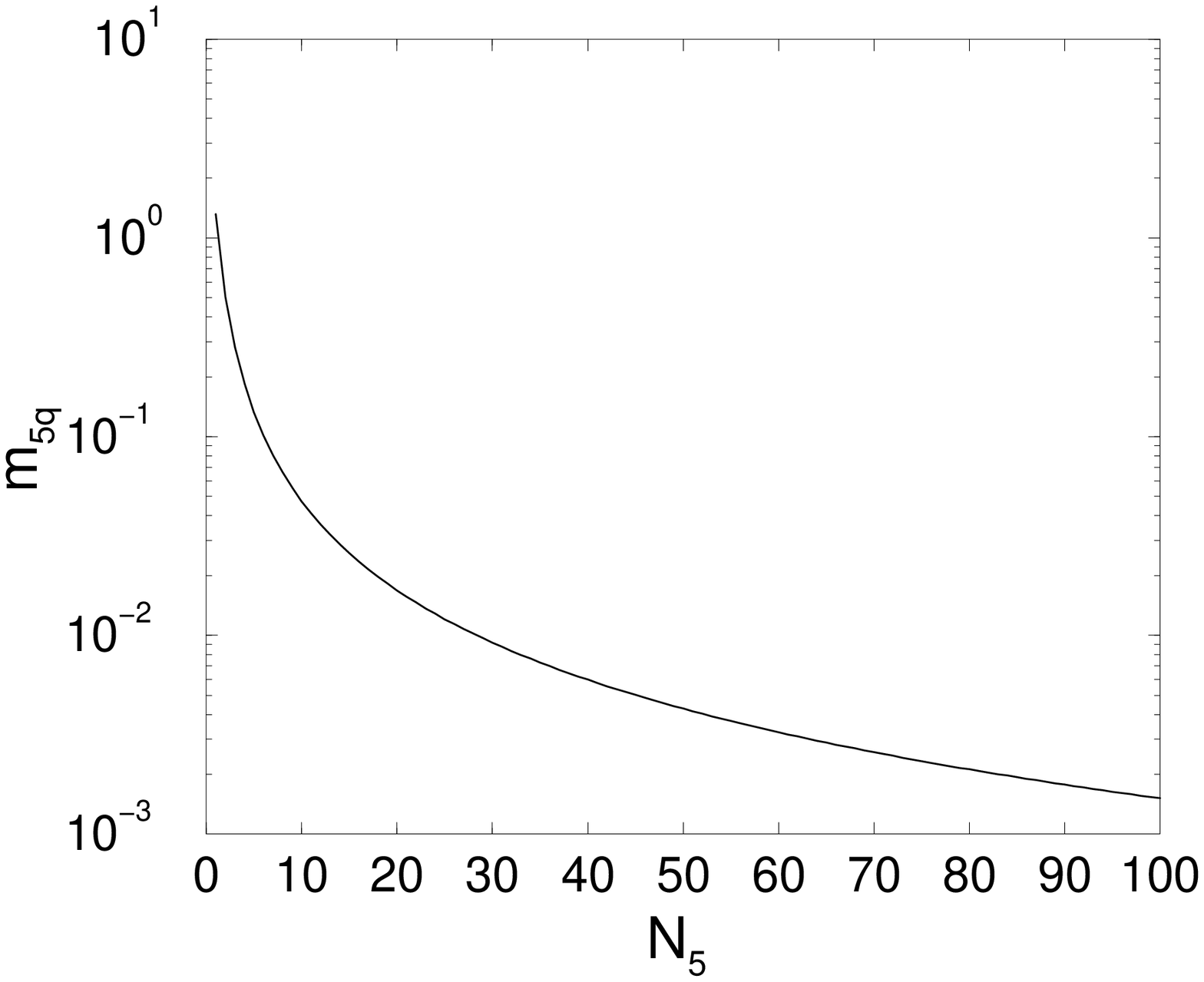}
  \epsfxsize=5.5cm \epsfbox{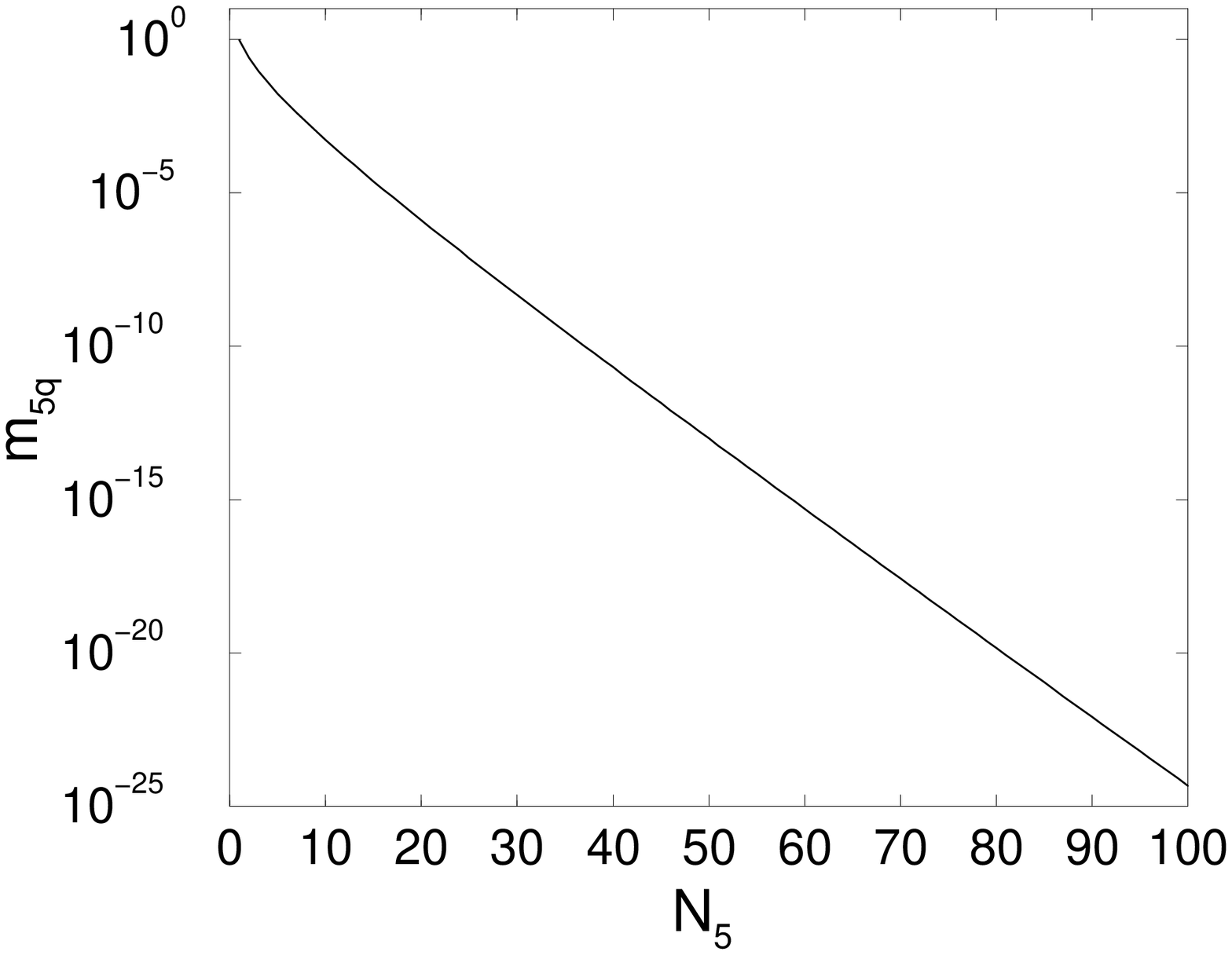}
\caption{$m_{5q}$ as function of $N_5$ evaluated with three types of
  $\rho(\lambda)$. The left one corresponds to type (1) with
  $\delta=-0.5$, the center is type (2) with $\delta=0$ and the right is
  type (3) with $\delta=0.5$.} 
\label{fig:m5q}
 \end{center}
\end{figure}

For $\delta<0$ the asymptotic behavior of the anomalous quark mass
becomes 
\begin{eqnarray}
m_{5q} &\to&
\frac{\rho(0)}{2N_5}+{\cal O}\left(\frac{1}{N_5^2}\right).
\end{eqnarray}
For $\delta=0$
\begin{eqnarray}
m_{5q} &\to&
\sqrt{\frac{\pi R}{2}}(1-\sqrt{2})\zeta\left(\frac{1}{2}\right)
 \frac{1}{N_5^{3/2}}
+{\cal O}\left(\frac{1}{N_5^{5/2}}\right),
\end{eqnarray}
where $\zeta$ is the Riemann's zeta function.
For $\delta>0$ we have
\begin{eqnarray}
m_{5q} &\to&
e^{-\delta N_5}\left(
\sqrt{\frac{\pi R}{2}}\frac{1}{N_5^{3/2}}
+{\cal O}\left(\frac{1}{N_5^{5/2}}\right)\right).
\end{eqnarray}
We can see that $\delta=0$ gives power law even if the zero mode density 
vanishes $\rho(0)=0$.
We need a gap at $\lambda =0$ in the continuous eigenvalues
to realize exponential decay.
Note that the similar analysis is also made in Ref.~\cite{Shamir00}.
Typical form of $m_{5q}$ is given in Fig.~\ref{fig:m5q} for $R=4$ and
$\delta=-0.5$, $\delta=0$, $\delta=0.5$.

\section{Effect of localized eigenstates and the infinite volume limit
 of the system}
\label{sec:analysis}
In this section, using the formula for $m_{5q}$ in terms of the eigenvalues,
we propose our interpretation, which resolves the inconsistency
among the numerical simulations for the domain-wall QCD, mentioned in the 
introduction.

We consider the situation that the density of the continuous eigenvalues
has a gap at zero: $\rho(\lambda) = 0$ for $\vert \lambda \vert \le \delta $ 
with $\delta > 0$.
On the other hand, we do not put such a restriction on
the localized eigenvalues, so that they can become almost zero.
In this situation, the contribution of the continuous eigenvalues to $m_{5q}$
vanishes exponentially in $N_5$.
In the following subsections we will consider the contribution from the 
localized eigenstates to $m_{5q}$.

\subsection{$N_5\rightarrow \infty$ at finite volume}

In this subsection we discuss the behavior of $m_{5q}$ in the large $N_5$
at finite volume. This situation often corresponds to the one encountered in 
the numerical simulations, and in the infinite $N_5$ limit the domain-wall 
fermion becomes the overlap Dirac operator, which satisfies 
the Ginsparg-Wilson relation.

In the finite volume, it is almost impossible to have an {\it exact} 
zero eigenvalue.
More precisely a probability to have the exact zero eigenvalue is zero,
since no symmetry assures the existence of it.
Indeed no exact zero is numerically found in the evaluations of
eigenvalues of $H_W$.
On the other hand, {\it almost} zero eigenvalues, $\vert \lambda \vert 
\simeq 10^{-2}$ or less, appear, and they become smaller as the volume 
increases. Therefore it may be reasonable to assume that
the average of the smallest eigenvalue vanishes in some power of $1/V$:
$\langle \vert \lambda_{\rm min}\vert \rangle = c_0 V^{-c_1}$ with $c_0, c_1
> 0$.

In this situation $m_{5q}$ vanishes exponentially as $N_5$ increase:
\begin{eqnarray}
m_{5q} &\propto& \frac{1}{4 N_c V} \exp [ - \frac{ N_5 c_0}{ V^{c_1}} ]
\end{eqnarray}
as $N_5\rightarrow\infty$. This means that DWQCD always works in the finite 
volume: $m_{5q}$ vanishes exponentially in $N_5$.
In other words, as long as the continuous eigenvalues have a gap around zero,
DWF in the $N_5\rightarrow\infty$ limit, or equivalently,
the overlap Dirac fermion, works well to describe the chiral modes in the
finite volume, where
the smallest eigenvalue of the localized modes is small but non-zero.

According to this consideration, we speculate how DWQCD behaves as the coupling
constant varies. 
In the strong coupling, even the continuous eigenvalues have no gap such that
$\vert\lambda_{\rm min}\vert = 0$, and therefore
$m_{5q}$ vanishes only in some power of $1/N_5$. DWQCD does not work in the
strong coupling region. 
Once the gap in the continuous eigenvalues opens ($\vert \lambda_{\rm min}
\vert > \delta > 0$) at $\beta > \beta_c$ in the weak coupling region,
$m_{5q}$ vanishes exponentially in $N_5$ and DWQCD in the finite volume
works well to describe the chiral symmetry.

We think that $\beta_c < 6.0$ for the plaquette action and $\beta_c < 2.6$
for the RG improved action. At first sight this seems to contradict with 
the numerical data of $m_{5q}$, which does not show the exponential decay in 
$N_5$, for the plaquette action\cite{cppacs-dwf,RBC}.  
This contradicted behavior is explained as follows.
In the intermediate values of $N_5$, the continuous eigenvalues give the
main contribution of $m_{5q}$, so that 
\begin{eqnarray}
m_{5q} &\sim& C_{\rm cont}\exp [-N_5\vert\lambda_{\rm min}^{\rm cont}\vert],
\end{eqnarray}
while as $N_5$ further increases the localized eigenvalues dominates, so that
\begin{eqnarray}
m_{5q} &\sim& C_{\rm local}\exp [-N_5\vert\lambda_{\rm min}^{\rm local}\vert] ,
\end{eqnarray}
where $C_{\rm cont}$ or $C_{\rm local}$ is proportional to the number of
the modes near $\lambda_{\rm min}^{\rm cont}$ or 
$\lambda_{\rm min}^{\rm local}$, respectively.
At $\beta=6.0$ for the plaquette action, the transition
between the former exponential behavior with 
$\vert\lambda_{\rm min}^{\rm cont}\vert$ and the latter with
$\vert\lambda_{\rm min}^{\rm local}\vert$ can be seen for $N_5 = 10\sim 40$
\cite{cppacs-dwf,RBC}.
The latter behavior looks almost constant in $N_5$ since 
$\lambda_{\rm min}^{\rm local}$ is very small.
At $\beta=2.6$ for the RG improved action, on the other hand,
only the former exponential behavior can be detected for $ N_5 = 10\sim 24$
\cite{cppacs-dwf}.
This suggests that the ratio $C_{\rm local}/C_{\rm cont}$ is smaller for the RG
action than for the plaquette action. 
Indeed it has been numerically found\cite{cppacs-nagai} 
that the number of the localized modes
near $\lambda_{\rm min}^{\rm local}$ is much less for the RG action, while
the number of the continuous modes near $\lambda_{\rm min}^{\rm cont}$
is similar in the two actions. (However $\vert\lambda_{\rm min}^{\rm local}
\vert$ seems larger for the RG action.)
We then give two predictions, which should be checked in order to test
this interpretation:
the exponential decay,
$\exp [-N_5\vert\lambda_{\rm min}^{\rm local}\vert] $, can be seen
at larger $N_5 > 60$ for the plaquette action, and the transition between
the former and the latter can be seen at $N_5 = 40\sim 60$ for the RG action.

The distribution of eigenvalues of $H_W$\cite{cppacs-nagai} suggests that
even $\beta= 5.65$ for the plaquette action or $\beta=2.2$ for the
RG action is already in the weak coupling region ($\beta > \beta_c$).
Indeed it seems that the transition between two exponentials has been observed 
in the behavior of $m_{5q}$ at these $\beta$\cite{cppacs-dwf}.

\subsection{Infinite volume limit at finite $N_5$}

In the previous subsection, we argue that $m_{5q}$ decay exponentially
in $N_5$ in the finite volume as long as the distribution of
the continuous eigenvalues has a gap around zero.
The exponential decay rate $\vert \lambda_{\rm min}^{\rm local}\vert$,
however, may vanish in the infinite volume limit.
Therefore, it may be the case that $m_{5q}$ does not vanish exponentially
if the infinite volume limit is taken before $N_5\rightarrow\infty$.
Moreover, the value of $N_5$ necessary for the suppression of the
chiral symmetry breaking, $m_{5q}$, may increase as the volume
becomes larger. This may be a disaster from the practical point of view.
In this subsection, we will discuss whether the effect of the localized 
eigenstates to $m_{5q}$ vanishes or not
in the infinite volume limit.

In the case of the hermitian Wilson-Dirac operator, 
the plane-wave eigenvalues in the finite volume 
become the continuous spectra in the infinite volume. 
On the other hand, we can not predict the nature of the distribution
for the localized eigenvalues in the infinite volume limit.
Although the localized modes are always discrete in the finite volume,
it can be continuous in the limit as will be discussed later.

We first consider the case where the number of the localized modes around
zero does not grow linear in the volume.
More precisely, the number of modes which satisfy $-\epsilon< \lambda < 
\epsilon$ with $\epsilon > 0$, denoted by $N(-\epsilon, \epsilon)$,
is bounded by $ c_0 \epsilon V^{c_1}$ with $c_0 >0$ and $c_1 < 1$.
In this case the contribution to $m_{5q}$ in the infinite volume limit 
becomes
\begin{eqnarray}
m_{5q}^{\rm localized}
&\simeq& \lim_{V\rightarrow\infty}
\frac{1}{V} \sum_{n:{\rm localized}} e^{- \vert\lambda_n\vert N_5}
< \lim_{V\rightarrow\infty}
\frac{N(-\epsilon,\epsilon)}{V} + O(e^{-\epsilon N_5})\nn\\
&<& \lim_{V\rightarrow\infty}
c_0 \epsilon V^{c_1-1} + O(e^{-\epsilon N_5})
= O(e^{-\epsilon N_5}),
\end{eqnarray}
where the sum is taken only for the localized modes.
This result shows that the contribution from the localized near-zero modes
vanishes in the infinite volume limit and the remaining contribution
also vanishes exponentially in $N_5$,
as long as the number of the localized near-zero modes is bounded
as
$ N(-\epsilon,\epsilon)= c_0 \epsilon V^{c_1}$ with $c_1 < 1$.
DWQCD works well in this case even in the infinite volume limit.

If the number of the localized near zero-modes is proportional to
the volume; $N(-\epsilon,\epsilon) = c_0 \epsilon V$, 
the contribution from them
remains non-zero in the infinite volume limit:
\begin{eqnarray}
m_{5q}^{\rm localized}
\lim_{V\rightarrow\infty}
\simeq \frac{1}{V} \sum_{n:{\rm localized}} e^{- \vert\lambda_n\vert N_5}
= c_0 \epsilon  + O(e^{-\epsilon N_5}) .
\end{eqnarray}
DWQCD does not work in this case in the infinite volume limit.

If $N(-\epsilon,\epsilon) = c_0 \epsilon V$,
the density of state $\rho(\lambda)$ is (almost) continuous
and nonzero at $\lambda = 0$ in the infinite volume limit:
$\rho(0)\not= 0$. In other words, a sum of the infinitely many
$\delta$ functions from the localized modes, normalized by the volume, becomes 
the continuous function in the infinite volume limit. 
To be more precise, one should first define the integrated density of states
by
\begin{eqnarray}
k(\lambda) &=&
\lim_{V\rightarrow\infty}\frac{1}{V} N(-\infty, \lambda).
\end{eqnarray}
Since $k(\lambda)$ is a monotonically increasing function in $\lambda$,
its derivative, $\displaystyle\frac{d k(\lambda)}{d \lambda}$, becomes
a well-defined measure. The previous statement is equivalent to that
$\rho_{\rm localized}(\lambda)\equiv
\displaystyle\frac{d k(\lambda)}{d \lambda}$ is continuous and non-zero
at $\lambda = 0$.

Let us explain the case that $\rho_{\rm localized}(0)\not=0$ 
more concretely by using a model of the
eigenvalue distribution. Suppose that the discrete eigenvalues are uniformly
distributed in the interval that $ -1 < \lambda < 1$ and the number of
modes is equal to $ 2 c V$, so that the average interval between two
successive eigenvalues becomes $ 1/(c V)$. In the infinite volume limit,
we see that the eigenvalues are discrete but dense in the interval.
If we calculate $m_{5q}$ using this distribution, we have
\begin{eqnarray}
m_{5q} &=&
%\lim_{V\rightarrow\infty}\frac{1}{V}\sum_{n=-cV}^{cV} 
%e^{-\frac{\vert n\vert }{cV} N_5}=
\lim_{V\rightarrow\infty}\frac{2}{V}\sum_{n=1}^{cV}
e^{-\frac{ n }{cV} N_5}
%\nn\\&=&
=\lim_{V\rightarrow\infty}\frac{2}{V}\frac{e^{-\frac{N_5}{cV}}
(1-e^{-N_5})}{1-e^{-\frac{N_5}{cV}}}
= \frac{2c}{N_5}(1-e^{-N_5}) .
\end{eqnarray}
Therefore $m_{5q}$ does not vanish exponentially in $N_5$ if the infinite
volume limit is first taken.

The appearance of the continuous density of states from
the localized modes in the infinite volume
is often observed
if a kind of randomness is introduced in the 
interaction\cite{neuberger,luscher,nakamura}.
Physically the Anderson localization\cite{anderson} is one of such examples.
In QCD one can have infinitely many localized modes
if infinitely many pairs of instanton and 
anti-instanton with a fixed topological charge exist.
In lattice QCD, we have, in addition, very localized modes associated
with the dislocations. Since the dislocations is local,
the number of the discrete modes associated with them can 
be proportional to the volume\cite{BNN,luscher}.

Although the contribution of the localized modes to $m_{5q}$ vanishes 
in the finite volume,
whether it vanishes in the infinite volume limit
depends on the density of states of the localized modes in the
infinite volume limit.
The previous numerical investigations on $\rho(0)$\cite{EHN99,cppacs-nagai}
suggests $\rho(0)$ is small but non-zero at $\forall\beta\not=\infty$
in quenched QCD.
Further investigations, however, are necessary in particular for the volume
dependence of $\rho(0)$, to have a definite conclusion on this point
whether $\rho(0)$ is non-zero or not.
If $\rho(0)\not=0$ at $\forall\beta\not=\infty$, it is interesting and
important to derive the $\beta$ dependence of $\rho(0)$\cite{EHN99}
theoretically.
Note however that $\rho(0)$ should vanish in the continuum limit for
$ 0 < M < 2$, since no zero mode appears in the free theory.

\subsection{Phase structure with the Wilson quarks}

In this subsection we consider the relation between the above interpretation
for the behavior of $m_{5q}$ in DWQCD and the parity-flavor breaking in the
Wilson quark action in the quenched approximation.

We first show that the contribution of the localized modes to the parity-flavor
breaking order parameter vanishes in the infinite volume limit,
in the case that $N(-\epsilon,\epsilon) \le c_0 \epsilon V^{c_1}$ 
with $c_1 < 1$.
In this case we have
\begin{eqnarray}
\langle \bar\psi i\gamma_5\tau^3 \psi \rangle^{\rm localized} &=&
\lim_{H\rightarrow 0}\lim_{V\rightarrow\infty}\frac{1}{V}
\sum_{n:{\rm localized}} \frac{-2H}{\lambda_n^2 + H^2} \nn\\
&\le & \lim_{H\rightarrow 0}\lim_{V\rightarrow\infty}\frac{1}{V}
\left[ \epsilon c_0 V^{c_1} \frac{-2 H}{H^2} 
+\sum_{n: \vert\lambda_n \vert >\epsilon}\frac{-2H}{\epsilon^2 + H^2} \right]
\nn\\
&\le&  \lim_{H\rightarrow 0}\lim_{V\rightarrow\infty}c_0\epsilon V^{c_1-1}
\frac{-2}{H}
+ \lim_{H\rightarrow 0}\lim_{V\rightarrow\infty}\frac{1}{V}
C(\epsilon) V \frac{-2H}{\epsilon^2 + H^2} 
\nn\\
&\le & C(\epsilon) \lim_{H\rightarrow 0}\frac{-2H}{\epsilon^2 + H^2} = 0,
\end{eqnarray}
where $\sum_{n: \vert\lambda_n \vert >\epsilon} 1 \equiv C(\epsilon) V$.
The first term vanishes in the $V\rightarrow\infty$ limit, while
the second one vanishes in the $H\rightarrow 0$ limit.

We next show that the contribution of the localized modes to the parity-flavor
breaking order parameter remains non-zero in the infinite volume limit,
if the density of such states is non-zero at $\lambda =0$ in the
infinite volume limit: $N(-\epsilon,\epsilon) = c_0 \epsilon V$.
To see this, we again use the uniform distribution of eigenvalues
between $-1$ and $1$, used in the previous subsection.
\begin{eqnarray}
\langle \bar\psi i\gamma_5\tau^3 \psi \rangle^{\rm localized} &=&
\lim_{H\rightarrow 0}\lim_{V\rightarrow\infty}\frac{2}{V}
\sum_{n=1}^{cV} \frac{-2H}{(\frac{n}{cV})^2 + H^2} 
=  -4c\lim_{H\rightarrow 0}\lim_{V\rightarrow\infty}
\sum_{n=1}^{cV}\frac{HcV}{n^2 + (HcV)^2} \nn\\
&=&  -4c\lim_{H\rightarrow 0}\lim_{V\rightarrow\infty}
\left[ \frac{\pi {\rm coth}(\pi HcV)}{2}-\frac{1}{2HcV} + \Delta
\right]
=-2c\pi,
\end{eqnarray}
where $\lim_{H\rightarrow 0}\lim_{V\rightarrow\infty} \Delta \le 
\lim_{H\rightarrow 0} \pi H/(2\sqrt{1+H^2}) = 0$.
 
The above two considerations show that the correspondence between
the failure of DWQCD in the infinite volume limit and 
non-zero order parameter of the parity-flavor breaking in the quenched 
Wilson fermion still holds even in the case that the localized modes
dominate in near-zero eigenvalues.
The previous investigations\cite{EHN99,cppacs-nagai} suggests
that
the gap in the phase structure in Fig.~\ref{fig:phase} closes
even at $\beta > \beta_c$ in the quenched QCD with the Wilson quark.
In this case we expect that
non-zero value of $\rho_{\rm localized}(0)$ in the region A
is much smaller than $\rho_{\rm continuous}(0)$ in the region B,
the true parity-flavor breaking phase\cite{EHN99}, so that
$\rho(0)_{\rm localized}$ may be too small to be detected by
measuring the pion mass. This may be a reason why
the previous investigation indicates that 
the region $A$ exists at $\beta = 6.0$ for the plaquette 
action\cite{AKU}.
Finally it is noted that the region A with $\rho_{\rm localized}(0)=0$
should always exist in full QCD, 
since appearances of the localized near-zero modes are suppressed 
by the fermion determinant, $\det D_W = \det H_W$.

\subsection{Exceptional configurations}

Finally let us comment on the exceptional configurations appeared in the
quenched QCD with the Wilson-type quark action.
It is sometimes observed at small quark masses in the quenched simulations that
the pion propagator receives the anomalously large contribution from
a particular configuration, which is called the exceptional configuration
and is removed from the statistical average.
In ref.~\cite{AKU} the several configurations, which give the W shape
in the pion correlation function in $t$ space, have been observed in
the region A, where the gap(absence) of the party-flavor breaking phase 
is expected.
It has been very difficult to understand the W shape propagator, since
it means that the correlation increase as the separation $t$ increases.
Now we argue that this may be understood by the localized modes.

The pion correlation function is defined by
\begin{eqnarray}
\langle \bar\psi i\gamma_5\tau^a\psi (x)\cdot \bar\psi i\gamma_5\tau^a\psi (y) 
\rangle &=&  \VEV{x}{\frac{1}{H_w}}{y} \VEV{y}{\frac{1}{H_w}}{x} \nn\\
&=& \sum_n \braket{x}{\lambda_n}\frac{1}{\lambda_n}\braket{\lambda_n}{y}
\sum_l \braket{y}{\lambda_l}\frac{1}{\lambda_l}\braket{\lambda_l}{x}\nn\\
&=&
\sum_{n,l}\frac{1}{\lambda_n\lambda_l} \phi_n(x)\phi_l(x)^\dagger
\phi_l(y)\phi_n(y)^\dagger,
\end{eqnarray}
where $\phi_n(x)=\braket{x}{\lambda_n}$ etc. 
If some localized eigenstate has a very small eigenvalue $\lambda_n$
on some configuration, the contribution of the mode becomes very large:
\begin{eqnarray}
\langle \bar\psi i\gamma_5\tau^a\psi (x)\cdot \bar\psi i\gamma_5\tau^a\psi (y) 
&=& \frac{1}{\lambda_n^2}\phi_n(x)\phi_n(x)^\dagger \phi_n(y)\phi_n(y)^\dagger
+ \mbox{other contributions} .
\end{eqnarray}
This is the exceptional configuration.
Furthermore suppose the eigenstate $\phi_n(\vec x, t)$ is localized at
$x = (\vec x_n, t_n)$.
Summing over $\vec{x}$ and $\vec{y}$, and taking $x_0=t$ and $y_0=0$,
the time correlation is dominated by
\begin{eqnarray}
C(t)&\equiv& \sum_{\vec x,\vec y} \frac{1}{\lambda_n^2}
\phi_n(\vec x,t)\phi_n(\vec x,t)^\dagger 
\phi_n(\vec y,0)\phi_n(\vec y,0)^\dagger,
\end{eqnarray}
and $C(t)$ has a peak around $t = t_n$.
Together with an enhance factor $1/\lambda_n^2$ for small eigenvalues
this explains the W shape behavior of the pion propagator.

\reseteqnum
%%%%%%%%%%%%%%%%%%%%%%%%%%%%%Section 7%%%%%%%%%%%%%%%%%%%%%%%%%%%%%%%
\section{Conclusions}
\label{sec:conclusion}

In this paper we first derive the formula for the chiral symmetry
breaking term $m_{5q}$ in DWQCD in terms of the (modified) hermitian
Wilson-Dirac operator in 4 dimensions. 
Using several simplifications and approximations 
we explicitly write down the formula of $m_{5q}$ in terms the
eigenvalues only.

The important observation  is that there are two different
types for the eigenvalues of the
the hermitian Wilson-Dirac operator. One is the continuous one, which 
corresponds to the plane-wave in the free theory, 
the other is the localized one, associated with the instanton or dislocations.
We argue that the effect of the latter one to the physical observable
such as $m_{5q}$ or the parity-flavor breaking order parameter vanishes
in the infinite volume limit, unless the number of near zero localized
modes increases linearly in the volume.
On the other hand, the effect remains non-zero but small
in the infinite volume limit, if it linearly increases.

The message of this paper is that domain-wall fermion or 
overlap Dirac fermion should work well to describe the vector-like
chiral symmetry at weak coupling, $\beta > \beta_c$, where the gap in
the continuous spectra opens, as long as the volume is finite.
The small eigenvalues appeared in the numerical simulation seem to
belong to the localized one, hence their contribution vanishes
in the $N_5\rightarrow \infty$ limit in the finite volume.
Whether domain-wall/overlap fermions are successful or not in the 
infinite volume, however, depends crucially on the distribution 
of the localized eigenvalues in the limit.

In practice one can make such small eigenvalues appeared in the
finite volume larger by hand,
without changing physical observables\cite{EH00,HJL00}.
This is a little costly.
If the effect of chiral symmetry breaking is small enough or no
dependence of observables on $N_5$ is detected, one may instead
perform simulations at large but numerically affordable value 
of $N_5$. Except the quantities very sensitive to the small eigenvalues
such as $m_{5q}$, $N_5$ dependences are expected to be
rather week in general.
This expectation is indeed true in the case of the quantum
hall effect\cite{QHE}.

%%%%%%%%%%%%%%%%%%%%%%%%%%%%%Section  %%%%%%%%%%%%%%%%%%%%%%%%%%%%%%%
\section*{Acknowledgments}

We thank Profs. Y.~Kikukawa, T.~Izubuchi and K.~I.~Nagai
for useful discussion.
S.A would like to thank Profs. M.~L\"uscher, H.~Neuberger and
S.~Nakamura for informative discussion, and
Y.T would like to thank  Prof. T.~Onogi for valuable
discussion.
This work is supported in part by Grants-in-Aid of the Ministry of
Education (Nos. 12304011,12640253, 13135204).

\reseteqnum
%%%%%%%%%%%%%%%%%%%%%%%%%%%%%Section 7%%%%%%%%%%%%%%%%%%%%%%%%%%%%%%%
\appendix
\section{Plane wave}
\label{sec:appendix-a}

The eigenvector \eqn{eqn:plane-wave} of the free Hamiltonian
\eqn{eqn:free-H} is given by combining positive and negative energy
eigenvector $u_\alpha(p,\tau)$ and $v_\alpha(p,\tau)$
\begin{eqnarray}
&&
 (H')_{\alpha\beta} v_\beta(p,\tau)=-|\lambda'(p)|v_\alpha(p,\tau),
\\&&
 (H')_{\alpha\beta} u_\beta(p,\tau)=|\lambda'(p)|u_\alpha(p,\tau),
\end{eqnarray}
where $\alpha$, $\beta$ are spinor indexes and $\tau$ runs $\tau=1,2$.
We adopt the following combination in this paper;
\begin{eqnarray}
 U_\alpha(p,1)=v_\alpha(p,1),\;
 U_\alpha(p,2)=v_\alpha(p,2),\;
 U_\alpha(p,3)=u_\alpha(p,1),\;
 U_\alpha(p,4)=u_\alpha(p,2).
\end{eqnarray}
$u$ and $v$ are given as follows with Pauli matrix $\vec{\sigma}$
and two dimensional basis vector $\xi(\tau)$,
\begin{eqnarray}
&&
v_\alpha(p,\tau)=\sqrt{\frac{|\lambda'(p)|-B(p)}{2|\lambda'(p)|}}
 \pmatrix{\xi(\tau) \cr
 \frac{-i\sigma_\mu^\dagger A_\mu(p)}{B(p)-|\lambda'(p)|}\xi(\tau)}_\alpha,
\\&&
u_\alpha(p,\tau)=\sqrt{\frac{|\lambda'(p)|-B(p)}{2|\lambda'(p)|}}
 \pmatrix{\frac{-i\sigma_\mu A_\mu(p)}{B(p)-|\lambda'(p)|}\xi(\tau) \cr
 \xi(\tau)}_\alpha,
\\&&
\sigma_\mu=\left(1,-i\vec{\sigma}\right),
\\&&
\xi(1)=\pmatrix{1 \cr 0},\; \xi(2)=\pmatrix{0 \cr 1}.
\end{eqnarray}
$U_\alpha(p,s)$ satisfies the orthogonal and completeness condition
\begin{eqnarray}
&&
\sum_{\alpha}U_\alpha^\dagger(p,s)U_\alpha(p,t)=\delta_{s,t},
\\&&
\sum_{s}U_\alpha^\dagger(p,s)U_\beta(p,s)=\delta_{\alpha,\beta}.
\end{eqnarray}

\section{Anomalous quark mass with completely localized and plane-wave 
eigenvectors}
\label{sec:appendix-b}

We consider a system with $N_l$ completely localized eigenvectors and
plane-wave functions with $N_p$ degrees of freedom for momentum $p$.

We expand the two Green functions in $m_{5q}$ with
the complete set in eq.(\ref{eqn:complete-1}).
\begin{eqnarray}
X(t) &=& \sum_{\vec{x}}
\vev{J_{5q}(x)P(y)}
\nn\\&=&
-\frac{a_5}{a^{10}}\sum_{\vec{x},\alpha,a}\sum_{b,\beta}
\VEV{I}{\frac{1}{2\cosh\frac{N_5}{2}a_5\wt{H}}D_{N_5}^{-1}}{J}
\VEV{J}{\left(D_{N_5}^{-1}\right)^\dagger
\frac{1}{2\cosh\frac{N_5}{2}a_5\wt{H}}}{I}
\nn\\&=&
-\frac{a_5}{a^{10}}\sum_{\vec{x},\alpha,a}\sum_{b,\beta}
\Biggl(
\sum_{n,m=1}^{N_l}\braket{I}{n}
\VEV{n}{\frac{1}{2\cosh\frac{N_5}{2}a_5\wt{H}}D_{N_5}^{-1}}{J}
\nn\\&&\times
\VEV{J}
{\left(D_{N_5}^{-1}\right)^\dagger\frac{1}{2\cosh\frac{N_5}{2}a_5\wt{H}}}{m}
\braket{m}{I}
\nn\\&+&
\sum_{p,s,c}\sum_{k,t,d}\braket{I}{p,s,c}
\VEV{p,s,c}{\frac{1}{2\cosh\frac{N_5}{2}a_5\wt{H}}D_{N_5}^{-1}}{J}
\nn\\&&\times
\VEV{J}
{\left(D_{N_5}^{-1}\right)^\dagger\frac{1}{2\cosh\frac{N_5}{2}a_5\wt{H}}}
{k,t,d}
\braket{k,t,d}{I}
\Biggr),
\end{eqnarray}
where $I=(x,\alpha,a)$, $J=(y,\beta,b)$ and we use a relation
$\braket{I}{n}\braket{k,t,d}{I}=0$.

Contribution from the localized eigenvector is given as follows 
\begin{eqnarray}
X_l(t) &=&
-\frac{a_5}{a^{10}}\sum_{\vec{x},\alpha,a}\sum_{b,\beta}
\sum_{n,m=1}^{N_l}
\braket{I}{n}\braket{m}{I}
\nn\\&&\times
\VEV{n}{\frac{1}{2\cosh\frac{N_5}{2}a_5\wt{H}}D_{N_5}^{-1}}{J}
\VEV{J}
{\left(D_{N_5}^{-1}\right)^\dagger\frac{1}{2\cosh\frac{N_5}{2}a_5\wt{H}}}{m}
\nn\\&=&
-\frac{a_5}{a^{10}}\sum_{n=1}^{N_l}\delta_{t,x_n^0}
\left(\frac{1}{2\cosh\frac{N_5}{2}a_5\wt{\lambda}(\lambda'_n)}\right)^2
\sum_{b,\beta}
\VEV{n}{D_{N_5}^{-1}}{y,\beta,b}
\VEV{y,\beta,b}{\left(D_{N_5}^{-1}\right)^\dagger}{n},
\nn\\
\end{eqnarray}
where we use a relation
\begin{eqnarray}
&&
\braket{I}{n}=\braket{x,\alpha,a}{n}=
\delta_{\alpha,\alpha_n}\delta_{a,a_n}\delta_{x,x_n},
\\&&
\braket{I}{n}\braket{m}{I}
=\delta_{m,n}
\delta_{\alpha,\alpha_n}\delta_{a,a_n}\delta_{x,x_n}.
\end{eqnarray}
By inserting the identity operator in $(x,\alpha,a)$ space
\begin{eqnarray}
 1=\sum_{x,\alpha,a}\ket{x,\alpha,a}\bra{x,\alpha,a}
\end{eqnarray}
we have
\begin{eqnarray}
X_l(t) &=&
-\frac{a_5}{a^{10}}\sum_{n=1}^{N_l}\delta_{t,x_n^0}
\left(\frac{1}{2\cosh\frac{N_5}{2}a_5\wt{\lambda}(\lambda'_n)}\right)^2
\nn\\&&\times
\sum_{b,\beta}
\sum_{z,\gamma,c}\sum_{w,\delta,d}
\braket{n}{z,\gamma,c}
\VEV{z,\gamma,c}{D_{N_5}^{-1}}{y,\beta,b}
\VEV{y,\beta,b}{\left(D_{N_5}^{-1}\right)^\dagger}{w,\delta,d}
\braket{w,\delta,d}{n}
\nn\\&=&
-\frac{a_5}{a^{10}}\sum_{n=1}^{N_l}\delta_{t,x_n^0}
\left(\frac{1}{2\cosh\frac{N_5}{2}a_5\wt{\lambda}(\lambda'_n)}\right)^2
\nn\\&&\times
\sum_{b,\beta}
\VEV{x_n,\alpha_n,a_n}{D_{N_5}^{-1}}{y,\beta,b}
\VEV{y,\beta,b}{\left(D_{N_5}^{-1}\right)^\dagger}{x_n,\alpha_n,a_n}
\nn\\&=&
-\frac{a_5}{a^{10}}\frac{1}{4N_cN_x}\sum_{n=1}^{N_l}
\left(\frac{1}{2\cosh\frac{N_5}{2}a_5\wt{\lambda}(\lambda'_n)}\right)^2
\nn\\&&\times
\sum_{\vec{x},\alpha,a}
\sum_{b,\beta}
\VEV{\vec{x},t;\alpha,a}{D_{N_5}^{-1}}{y,\beta,b}
\VEV{y,\beta,b}{\left(D_{N_5}^{-1}\right)^\dagger}{\vec{x},t;\alpha,a}
\nn\\&=&
\frac{1}{a_5}Y(t)\frac{1}{4N_cN_x}\sum_{n=1}^{N_l}
\left(\frac{1}{2\cosh\frac{N_5}{2}a_5\wt{\lambda}(\lambda'_n)}\right)^2,
\end{eqnarray}
where we assume that the center of the localized solution
$(x_n,\alpha_n,a_n)$ is distributed uniformly in the $(x,\alpha,a)$
space and we can take average over $(x,\alpha,a)$ when enough number of
configurations are summed in the simulation.

For the contribution from the  continuous eigenvalues we take
$t\to\infty$ limit to proceed,
\begin{eqnarray}
X_c(t)|_{t\to\infty} &=&
-\frac{a_5}{a^{10}}\sum_{\vec{x},\alpha,a}\sum_{b,\beta}
\sum_{p,s}\sum_{k,s'}
\frac{1}{N_p}U_\alpha(p,s)U_\alpha^\dagger(k,t)e^{i(p-k)x}
\nn\\&&\times
\VEV{p,s,a}{\frac{1}{2\cosh\frac{N_5}{2}a_5\wt{H}}D_{N_5}^{-1}}{y,\beta,b}
\VEV{y,\beta,b}
{\left(D_{N_5}^{-1}\right)^\dagger\frac{1}{2\cosh\frac{N_5}{2}a_5\wt{H}}}
{k,s',a}
\nn\\&=&
-\frac{a_5}{a^{10}}n_{xyz}'\frac{1}{N_p}\sum_{p,s,a}
\left(\frac{1}{2\cosh\frac{N_5}{2}a_5\wt{\lambda}(p)}\right)^2
\nn\\&&\times
\sum_{\beta,b}\VEV{p,s,a}{D_{N_5}^{-1}}{y,\beta,b}
\VEV{y,\beta,b}{\left(D_{N_5}^{-1}\right)^\dagger}{p,s,a},
\end{eqnarray}
where we use following relations
\begin{eqnarray}
&&
\braket{x,\alpha,a}{p,s,b}=
\frac{1}{\sqrt{N_p}}U_\alpha(p,s)e^{ipx}\delta_{a,b},
\\&&
\sum_{\vec{x}}e^{i(\vec{p}-\vec{k})\vec{x}}
=n'_{xyz}\delta_{\vec{p},\vec{k}},
\\&&
\lim_{t\to\infty}e^{it(p^0-k^0)} = \delta_{p^0,k^0},
\\&&
\sum_\alpha U_\alpha^\dagger(p,s)U_\alpha(p,s')=\delta_{s,s'}.
\end{eqnarray}
$n'_{xyz}$ is a volume of $\wt{\cal S}$ in $xyz$ space.
By inserting the complete set \eqn{eqn:complete-1} of eigenstates we
have 
\begin{eqnarray}
X_c(t)|_{t\to\infty} &=&
-\frac{a_5}{a^{10}}n_{xyz}'\frac{1}{N_p}\sum_{p,s,a}
\left(\frac{1}{2\cosh\frac{N_5}{2}a_5\wt{\lambda}(p)}\right)^2
\nn\\&\times&
\sum_{\beta,b}\Biggl(
\sum_{n,m=1}^{N_l}\VEV{p,s,a}{D_{N_5}^{-1}}{n}
\braket{n}{y,\beta,b}\braket{y,\beta,b}{m}
\VEV{m}{\left(D_{N_5}^{-1}\right)^\dagger}{p,s,a}
\nn\\&&+
\sum_{k,s',c}\VEV{p,s,a}{D_{N_5}^{-1}}{k,s',c}
 \braket{k,s',c}{y,\beta,b}
\nn\\&&\quad\times
\sum_{q,s'',d}\braket{y,\beta,b}{q,s'',d}
\VEV{q,s'',d}{\left(D_{N_5}^{-1}\right)^\dagger}{p,s,a}
\Biggr)
\nn\\&=&
-\frac{a_5}{a^{10}}n_{xyz}'\frac{1}{N_p}\sum_{p}
\left(\frac{1}{2\cosh\frac{N_5}{2}a_5\wt{\lambda}(p)}\right)^2
\nn\\&&\times
\Biggl(
\sum_{a,s}\sum_{n=1}^{N_l}\delta_{y,x_n}
\VEV{p,s,a}{D_{N_5}^{-1}}{n}\VEV{n}{\left(D_{N_5}^{-1}\right)^\dagger}{p,s,a}
\nn\\&&+
\frac{N_c}{N_p}\sum_{s,s'}
2a\left(\frac{-i\gamma_\mu C_\mu(p)+E(p)}{C^2+E^2}\right)_{s,s'}
2a\left(\frac{i\gamma_\mu C_\mu(p)+E(p)}{C^2+E^2}\right)_{s',s}
\Biggr)
\nn\\&=&
-\frac{a_5}{a^{8}}n_{xyz}'\frac{1}{N_p}\sum_{p}
\left(\frac{1}{2\cosh\frac{N_5}{2}a_5\wt{\lambda}(p)}\right)^2
F(p),
\end{eqnarray}
where $F(p)$ is defined as
\begin{eqnarray}
F(p) &=& F_l(p)+F_c(p),
\\
F_l(p) &=& \sum_{n=1}^{N_l}\delta_{y,x_n}\sum_{s,a}\frac{1}{a^2}
\VEV{p,s,a}{D_{N_5}^{-1}}{n}\VEV{n}{\left(D_{N_5}^{-1}\right)^\dagger}{p,s,a},
\\
F_c(p) &=& \frac{4N_c}{N_p}\left(\frac{4}{C^2(p)+E^2(p)}\right).
\end{eqnarray}
We can show that $F_l(p)$ vanishes as follows.
We extend the orthogonality assumption of the eigenvectors
$\braket{n}{p,s,a}=0$ to the case with $\gamma_5$;
$\VEV{n}{\gamma_5}{p,s,a}=0$.
This is plausible since the eigenfunction $\psi_n(x)=\braket{x}{n}$ is
localized in space-time and the overlap is suppressed even if we
multiply $\gamma_5$.
Then by using the explicit form \eqn{eqn:DN5}, the matrix element
of the truncated overlap Dirac operator $D_{N_5}$ turns out to be block
diagonal, in which only the matrix elements $\VEV{n}{D_{N_5}}{m}$ and
$\VEV{p,s,a}{D_{N_5}}{k,t,b}$ are non-zero and the off-diagonal part
$\VEV{p,s,a}{D_{N_5}}{n}$ becomes zero.
This block diagonal property is kept even if we take inversion and the
off-diagonal part becomes zero
\begin{eqnarray}
\VEV{p,s,a}{D_{N_5}^{-1}}{n}=0.
\end{eqnarray}
It is noted that in the free theory the relation
$D_W^\dagger D_W=D_WD_W^\dagger$ is satisfied and
$\VEV{n}{D_{N_5}^{-1}}{m}=0$ is shown exactly for non-degenerate
$|\lambda'_n|\neq|\lambda'_m|$.

The pion propagator in the denominator of $m_{5q}$ is expanded similarly
with eigenstates as
\begin{eqnarray}
Y(t) &=& \sum_{\vec{x}}\vev{P(x)P(y)}
\nn\\&=&
-\frac{a_5^2}{a^{10}}\sum_{\vec{x},a,\alpha}\sum_{b,\beta}
\VEV{I}{D_{N_5}^{-1}}{J}\VEV{J}{(D_{N_5}^{-1})^\dagger}{I}
\nn\\&=&
-\frac{a_5^2}{a^{10}}\sum_{\vec{x},a,\alpha}\sum_{b,\beta}
\Biggl(
\sum_{n,m=1}^{N_l}
\braket{I}{n}\VEV{n}{D_{N_5}^{-1}}{J}
\VEV{J}{(D_{N_5}^{-1})^\dagger}{m}\braket{m}{I}
\nn\\&+&
\sum_{p,s,c}\sum_{k,s',d}
\braket{I}{p,s,c}\VEV{p,s,c}{D_{N_5}^{-1}}{J}
\VEV{J}{(D_{N_5}^{-1})^\dagger}{k,s',d}\braket{k,s',d}{I}
\Biggr).
\end{eqnarray}
A contribution from localized eigenvector is given by
\begin{eqnarray}
Y_l(t)&=&
-\frac{a_5^2}{a^{10}}\sum_{n=1}^{N_l}
\delta_{t,x^0_n}\sum_{b,\beta}
\VEV{n}{D_{N_5}^{-1}}{y,\beta,b}
\VEV{y,\beta,b}{(D_{N_5}^{-1})^\dagger}{n}
\nn\\&=&
-\frac{a_5^2}{a^{10}}\sum_{n=1}^{N_l}
\delta_{t,x^0_n}\sum_{b,\beta}
\sum_{z,\gamma,c}\braket{n}{z,\gamma,c}
\VEV{z,\gamma,c}{D_{N_5}^{-1}}{y,\beta,b}
\nn\\&&\times
\sum_{w,\delta,d}\VEV{y,\beta,b}{(D_{N_5}^{-1})^\dagger}{w,\delta,d}
\braket{w,\delta,d}{n}
\nn\\&=&
-\frac{a_5^2}{a^{10}}\sum_{n=1}^{N_l}
\delta_{t,x^0_n}\sum_{b,\beta}
\VEV{x_n,\alpha_n,a_n}{D_{N_5}^{-1}}{y,\beta,b}
\VEV{y,\beta,b}{(D_{N_5}^{-1})^\dagger}{x_n,\alpha_n,a_n}
\nn\\&=&
-\frac{a_5^2}{a^{10}}
\sum_{n=1}^{N_l}
\frac{1}{4N_cN_x}\sum_{\vec{x},\alpha,a}\sum_{\beta,b}
\VEV{\vec{x},t,\alpha,a}{D_{N_5}^{-1}}{y,\beta,b}
\VEV{y,\beta,b}{(D_{N_5}^{-1})^\dagger}{\vec{x},t,\alpha,a}
\nn\\&=&
\frac{N_l}{4N_cN_x}Y(t).
\end{eqnarray}
Taking $t\to\infty$ limit a contribution from the continuous sector
becomes
\begin{eqnarray}
Y_c(t)|_{t\to\infty} &=&
-\frac{a_5^2}{a^{10}}\sum_{\vec{x},a,\alpha}\sum_{b,\beta}
\sum_{p,s}\sum_{k,s'}
\frac{1}{N_p}U_\alpha(p,s)U_\alpha^\dagger(k,s')e^{i(p-k)x}
\nn\\&&\times
\VEV{p,s,a}{D_{N_5}^{-1}}{y,\beta,b}
\VEV{y,\beta,b}{(D_{N_5}^{-1})^\dagger}{k,s',a}
\nn\\&=&
-\frac{a_5^2}{a^{10}}n_{xyz}'
\frac{1}{N_p}\sum_{p,s,a}\sum_{b,\beta}
\VEV{p,s,a}{D_{N_5}^{-1}}{y,\beta,b}
\VEV{y,\beta,b}{(D_{N_5}^{-1})^\dagger}{p,s,a}
\nn\\&=&
-\frac{a_5^2}{a^{10}}n_{xyz}'
\frac{1}{N_p}\sum_{p,s,a}
\nn\\&\times&
\sum_{b,\beta}\Biggl(
\sum_{n,m=1}^{N_l}\VEV{p,s,a}{D_{N_5}^{-1}}{n}\braket{n}{y,\beta,b}
 \braket{y,\beta,b}{m}\VEV{m}{(D_{N_5}^{-1})^\dagger}{p,s,a}
\nn\\&&
+\sum_{k,s',c}\VEV{p,s,a}{D_{N_5}^{-1}}{k,s',c}
 \braket{k,s',c}{y,\beta,b}
\nn\\&&\quad\times
\sum_{q,s'',d}\braket{y,\beta,b}{q,s'',d}
 \VEV{q,s'',d}{(D_{N_5}^{-1})^\dagger}{p,s,a}
\Biggr)
\nn\\&=&
-\frac{a_5^2}{a^{10}}n_{xyz}'\frac{1}{N_p}
\sum_{p}F(p).
\end{eqnarray}
As a consequence we have a relation
\begin{eqnarray}
Y(t\to\infty)=
-\frac{4N_cN_x}{4N_cN_x-N_l}
\frac{a_5^2}{a^{10}}n_{xyz}'
\frac{1}{N_p}\sum_{p}F(p).
\end{eqnarray}

Now we need to know behavior of $F(p)$ as a function of $p$ to further
proceed.
We can directly see that the function $F(p)=F_c(p)$ is almost constant
for all the range of $p$ as in Fig.~\ref{fig:Fp}.
We assume that $F(p)$ is independent of $p$.

By adopting this assumption the anomalous quark mass becomes
\begin{eqnarray}
m_{5q}&=&\frac{\lim_{t\to\infty}X(t)}{\lim_{t\to\infty}Y(t)}
\nn\\&=&
\frac{1}{a_5}\frac{1}{4N_cN_x}\sum_{n}
\left(\frac{1}{2\cosh\frac{N_5}{2}a_5\wt{\lambda}(\lambda'_n)}\right)^2
\nn\\&&
-\frac{1}{Y(t\to\infty)}
\frac{a_5}{a^{10}}n_{xyz}'\frac{1}{N_p}\sum_{p}
\left(\frac{1}{2\cosh\frac{N_5}{2}a_5\wt{\lambda}(p)}\right)^2
F(p)
\nn\\&=&
\frac{1}{a_5}\frac{1}{4N_cN_x}\sum_{n}
\left(\frac{1}{2\cosh\frac{N_5}{2}a_5\wt{\lambda}(\lambda'_n)}\right)^2
\nn\\&&
+\frac{4N_cN_x-N_l}{4N_cN_x}
\frac{1}{a_5}
\frac{1}{\sum_{k}F(k)}
\sum_{p}
\left(\frac{1}{2\cosh\frac{N_5}{2}a_5\wt{\lambda}(p)}\right)^2
F(p)
\nn\\&=&
\frac{1}{a_5}\frac{1}{4N_cN_x}
\left(
\sum_{n}\left(\frac{1}{2\cosh\frac{N_5}{2}a_5\wt{\lambda}(\lambda'_n)}\right)^2
+4N_c\sum_{p}\left(\frac{1}{2\cosh\frac{N_5}{2}a_5\wt{\lambda}(p)}\right)^2
\right)
\nn\\&=&
\frac{1}{a_5}\frac{1}{4N_cN_x}
\left(
\sum_{n}
\left(\frac{1}{2\cosh\frac{N_5}{2}a_5\wt{\lambda}(\lambda'_n)}\right)^2
+
\sum_{n}\rho_n\left(\frac{1}{2\cosh\frac{N_5}{2}a_5\wt{\lambda}_n}\right)^2
\right),
\end{eqnarray}
where $\rho_n$ is a number of degeneracy in $\wt{\lambda}_n$,
\begin{eqnarray}
\rho_n=\sum_{p,\alpha,a}\delta_{\wt{\lambda}_n,\wt{\lambda}(p)}.
\end{eqnarray}

\section{Anomalous quark mass with partially localized and 
plane-wave eigenvectors}
\label{sec:appendix-c}

We consider a system with $N_l$ subspaces ${\cal S}_n$
($n=1,\cdots,N_l$), which is spanned by partially localized eigenvectors
$\psi_n^{(i)}$.
We assume that each subspace ${\cal S}_n$ has no overlap and its
volume is $h_n$.
Remaining subspace $\wt{\cal S}$ of the system is spanned by plane-wave
eigenvectors as in the previous subsection.
If we set the size of $\wt{\cal S}$ to be $4N_cN_p$ the total degrees of
freedom of the system becomes
\begin{eqnarray}
4N_cN_x=\sum_{n=1}^{N_l}h_n+4N_cN_p.
\end{eqnarray}

The partially localized eigenvector is given by
\begin{eqnarray}
&&
H'\ket{n,i}=\lambda_n^{(i)}\ket{n,i},
\\&&
\braket{I}{n,i}=\frac{1}{\sqrt{h_n}}\psi_n^{(i)}(I),
\end{eqnarray}
where the function $\psi_n^{(i)}(I)$ is non-zero within ${\cal S}_n$.
%\begin{eqnarray}
% \psi_n^{(i)\dagger}(I)\psi_n^{(i)}(I)=\cases{
%1 & for $I\in {\cal S}_n$ \cr
%0 & otherwise \cr}.
%\end{eqnarray}
We have a relation that
\begin{eqnarray}
\psi_n^{(i)\dagger}(I)\psi_m^{(j)}(I)=0\quad
\end{eqnarray}
for $n\neq m$ for any $i,j$.
The orthogonality is given by
\begin{eqnarray}
\sum_{I}\psi_n^{(i)\dagger}(I)\psi_m^{(j)}(I)
 =h_n \delta_{n,m}\delta_{i,j}.
\end{eqnarray}
The plane-wave eigenvector is the same as in the previous section
\eqn{eqn:plane-wave} with a constraint
\begin{eqnarray}
\braket{n,i}{I}\braket{I}{p,s,b}=0.
\end{eqnarray}
The completeness condition of this system is given by
\begin{eqnarray}
1=\sum_{n=1}^{N_l}\sum_{i=1}^{h_n}\ket{n,i}\bra{n,i}
+\sum_{l,s}\sum_{p}\sum_{a}\ket{p,l,s,a}\bra{p,l,s,a}
\label{eqn:complete-2}
\end{eqnarray}
with
\begin{eqnarray}
\sum_{n=1}^{N_l}\sum_{i=1}^{h_n}\frac{1}{h_n}\psi_n^{(i)}(I)
\psi_n^{(i)\dagger}(J)
=\delta_{I,J}.
\end{eqnarray}

We expand the two Green functions in $m_{5q}$ with the complete set in
the above.
\begin{eqnarray}
X(t) &=& \sum_{\vec{x}}\vev{J_{5q}(x)P(y)}
\nn\\&=&
-\frac{a_5}{a^{10}}\sum_{\vec{x},\alpha,a}\sum_{b,\beta}
\VEV{I}{\frac{1}{2\cosh\frac{N_5}{2}a_5\wt{H}}D_{N_5}^{-1}}{J}
\VEV{J}{\left(D_{N_5}^{-1}\right)^\dagger
\frac{1}{2\cosh\frac{N_5}{2}a_5\wt{H}}}{I}
\nn\\&=&
-\frac{a_5}{a^{10}}\sum_{\vec{x},\alpha,a}\sum_{b,\beta}\Biggl(
\sum_{n,m=1}^{N_l}\sum_{i,j=1}^{h_n}
\braket{x,\alpha,a}{n,i}
\VEV{n,i}{\frac{1}{2\cosh\frac{N_5}{2}a_5\wt{H}}D_{N_5}^{-1}}{J}
\nn\\&&\times
\VEV{J}
{\left(D_{N_5}^{-1}\right)^\dagger\frac{1}{2\cosh\frac{N_5}{2}a_5\wt{H}}}{m,j}
\braket{m,j}{x,\alpha,a}
\nn\\&+&
\sum_{p,s,c}\sum_{k,t,d}
\braket{x,\alpha,a}{p,s,c}
\VEV{p,s,c}{\frac{1}{2\cosh\frac{N_5}{2}a_5\wt{H}}D_{N_5}^{-1}}{J}
\nn\\&&\times
\VEV{J}
{\left(D_{N_5}^{-1}\right)^\dagger\frac{1}{2\cosh\frac{N_5}{2}a_5\wt{H}}}{k,t,d}
\braket{k,t,d}{x,\alpha,a}
\Biggr).
\end{eqnarray}
The contribution $X_c$ from the continuous eigenvalues is the same as in
the previous subsection except for the definition of $F(p)$
\begin{eqnarray}
F(p)&=& F_l(p)+F_c(p),
\\
F_l(p) &=&
\sum_{n=1}^{N_l}\sum_{i,j=1}^{h_n}\sum_{p,s,a}
\VEV{p,s,a}{D_{N_5}^{-1}}{n,i}
\frac{1}{h_n}\psi_n^{(i)\dagger}(J)\psi_n^{(j)}(J)
\VEV{n,j}{\left(D_{N_5}^{-1}\right)^\dagger}{p,s,a},
\\
F_c(p)&=&
\frac{4N_c}{N_p}\left(\frac{4}{C^2(p)+E^2(p)}\right).
\end{eqnarray}
Here we use the block diagonal condition
\begin{eqnarray}
\VEV{p,s,a}{D_{N_5}^{-1}}{n,i}=0
\end{eqnarray}
by assuming the $\gamma_5$ orthogonality $\VEV{n,i}{\gamma_5}{p,s,a}=0$.
$X_c$ is written in terms of the function $F(p)=F_c(p)$.

Contribution from the localized mode becomes
\begin{eqnarray}
X_l(t) &=&
-\frac{a_5}{a^{10}}
\sum_{b,\beta}
\sum_{n=1}^{N_l}\sum_{i,j=1}^{h_n}
\frac{1}{h_n}f_n^{i,j}(t)
\nn\\&&\times
\VEV{n,i}{\frac{1}{2\cosh\frac{N_5}{2}a_5\wt{H}}D_{N_5}^{-1}}{J}
\VEV{J}
{\left(D_{N_5}^{-1}\right)^\dagger\frac{1}{2\cosh\frac{N_5}{2}a_5\wt{H}}}{n,j}
\nn\\&=&
-\frac{a_5}{a^{10}}\sum_{n=1}^{N_l}\sum_{i,j=1}^{h_n}\frac{1}{h_n}f_n^{i,j}(t)
\left(\frac{1}{2\cosh\frac{N_5}{2}a_5\wt{\lambda}(\lambda^{'(i)}_n)}\right)
\left(\frac{1}{2\cosh\frac{N_5}{2}a_5\wt{\lambda}(\lambda^{'(j)}_n)}\right)
\nn\\&&\times
\sum_{b,\beta}
\sum_{z,\gamma,c}\sum_{w,\delta,d}
\braket{n,i}{z,\gamma,c}
\VEV{z,\gamma,c}{D_{N_5}^{-1}}{J}
\VEV{J}{\left(D_{N_5}^{-1}\right)^\dagger}{w,\delta,d}
\braket{w,\delta,d}{n,j}
\nn\\&=&
-\frac{a_5}{a^{10}}
\sum_{n=1}^{N_l}\sum_{i,j=1}^{h_n}\frac{1}{h_n}f_n^{i,j}(t)
\left(\frac{1}{2\cosh\frac{N_5}{2}a_5\wt{\lambda}(\lambda^{'(i)}_n)}\right)
\left(\frac{1}{2\cosh\frac{N_5}{2}a_5\wt{\lambda}(\lambda^{'(j)}_n)}\right)
\nn\\&&\times
\sum_{b,\beta}
\sum_{z,\gamma,c}\sum_{w,\delta,d}\frac{1}{h_n}
\psi_n^{(i)\dagger}(z,\gamma,c)
\VEV{z,\gamma,c}{D_{N_5}^{-1}}{J}
\VEV{J}{\left(D_{N_5}^{-1}\right)^\dagger}{w,\delta,d}
\psi_n^{(j)}(w,\delta,d),
\nn\\
\end{eqnarray}
where we assumed orthogonality for $m$ and $n$, $f_n^{i,j}$ is defined as
\begin{eqnarray}
\sum_{\vec{x}}\sum_{\alpha,a}\psi_n^{(i)}(I)\psi_m^{(j)\dagger}(I)
=\delta_{n,m}f_n^{i,j}(t).
\end{eqnarray}
$f_n^{i,j}(t)$ is non-zero when $t\in{\cal S}_n$.
Here we have a comment.
It is plausible to assume that the eigenfunction $\psi_n^{(i)}$
tends to be plane but rapidly oscillating inside ${\cal S}_n$ for large
$i$.
By using this fact we can see that the above $f_n^{i,j}$ is suppressed
for different $i$, $j$ even if the summation over $t$ is not taken.

We need more information on the propagator multiplied with eigenvectors
\begin{eqnarray}
\frac{1}{h_n}\sum_{I}\sum_{K}
\psi_n^{(i)\dagger}(I)
\VEV{I}{D_{N_5}^{-1}}{J}
\VEV{J}{\left(D_{N_5}^{-1}\right)^\dagger}{K}
\psi_n^{(j)}(K)
\label{eqn:propXeigen}
\end{eqnarray}
to further discuss the detailed property of $X_l$.
We start by adopting an assumption that the propagator
\begin{eqnarray}
f(I,K)=
\VEV{I}{D_{N_5}^{-1}}{J} \VEV{J}{\left(D_{N_5}^{-1}\right)^\dagger}{K}
\end{eqnarray}
is a slowly varying function of $I$ and $K$.
%, and we approximate it with a constant.
Then we investigate the behavior of \eqn{eqn:propXeigen} for two typical
forms of eigenfunctions.
The eigenfunctions are classified into two types,
according to  the behavior of
the following integral of a single function
\begin{eqnarray}
C=\sum_{I}\psi_n^{(i)}(I).
\end{eqnarray}

(i) Eigenfunction $\psi_n^{(i)}$ with $C\neq0$, which may be typical for
the lowest mode $i=1$.
It seems to be plausible \cite{HJL98,Nagai00} to approximate this lowest 
mode with the exponential form.
For simplicity we consider a one-dimensional case
\begin{eqnarray}
 \psi(x)=\frac{1}{\sqrt{\delta}}e^{-|x-x_0|/\delta},
\end{eqnarray}
where $\delta$ is a width and $x_0$ is a center of the eigenfunction.
Since the integral $\int dx\psi(x)$ gives non-zero value, a
multiplication with a
smooth function $f(x,y)$ produces a enhance factor $4\delta$, which is
proportional to a width of the eigenfunction:
\begin{eqnarray}
&&
\frac{1}{\delta}\int_{-\infty}^{\infty}dx dy e^{-|x-x_0|/\delta}f(x,y)
e^{-|y-x_0|/\delta}
\nn\\&&=
\frac{1}{\delta}\int_{-\infty}^{\infty}dz dw e^{-|z|/\delta}
f(z+x_0,w+x_0)e^{-|w|/\delta}
\nn\\&&=
\frac{1}{\delta}\int_{-\infty}^{\infty}dz dw e^{-|z|/\delta}
 f(x_0,x_0)e^{-|w|/\delta}
+{\cal O}(\p f(x_0,x_0))
=(4\delta)\ f(x_0,x_0).
\end{eqnarray}

The extension to four-dimensions  is given by
\begin{eqnarray}
 \psi_n^{(1)}(x)=\frac{1}{\delta^2}e^{-\sum_i|x_i-(x_n)_i|/\delta}
\end{eqnarray}
and \eqn{eqn:propXeigen} becomes
\begin{eqnarray}
&&
\frac{1}{h_n}\sum_{I}\sum_{K}
\psi_n^{(i)\dagger}(I)\VEV{I}{D_{N_5}^{-1}}{J}
\VEV{J}{\left(D_{N_5}^{-1}\right)^\dagger}{K}\psi_n^{(j)}(K)
\nn\\&&
=4^4h_n\delta_{i,j}
\VEV{I_n}{D_{N_5}^{-1}}{J}\VEV{J}{\left(D_{N_5}^{-1}\right)^\dagger}{I_n},
\end{eqnarray}
where $I_n$ is a peak of the exponential in $\psi_n^{(i)}(I)$ and $h_n$ is
given by $h_n=\delta^4$.
Here we use a property that the eigenfunction $\psi_n^{(i)}$ with
higher $i>1$ tends to be oscillating and a single summation
$\sum_{I}\psi_n^{(i)}(I)$ is suppressed.
Together with the suppression of $f_n^{i,j}$ for $i \neq j$ we have a
factor $\delta_{i,j}$ in the above.

The term $X_l$ of the numerator for the contribution
from the lowest eigenvector $i=1$ becomes 
\begin{eqnarray}
X_l(t) &=&
-\frac{a_5}{a^{10}}\sum_{n=1}^{N_l}\sum_{i=1}^{h_n}\delta_{i,1}4^4f_n^{i,i}(t)
\left(\frac{1}{2\cosh\frac{N_5}{2}a_5\wt{\lambda}(\lambda^{'(i)}_n)}\right)^2
\sum_{b,\beta}
\VEV{I_n}{D_{N_5}^{-1}}{J}
\VEV{J}{\left(D_{N_5}^{-1}\right)^\dagger}{I_n}
\nn\\&=&
-\frac{a_5}{a^{10}}\frac{1}{4N_cn_{xyz}}
\sum_{n=1}^{N_l}
%\sum_{i=1}^{h_n}
\frac{4^4h_n}{h_{n,t}}\delta_{t\in{\cal S}_n}
\left(\frac{1}{2\cosh\frac{N_5}{2}a_5\wt{\lambda}(\lambda^{'(1)}_n)}\right)^2
\nn\\&&\times
\sum_{\vec{x}}\sum_{\alpha,a}
\sum_{b,\beta}
\VEV{\vec{x},t,\alpha,a}{D_{N_5}^{-1}}{y,\beta,b}
\VEV{y,\beta,b}{\left(D_{N_5}^{-1}\right)^\dagger}{\vec{x},t,\alpha,a},
\nn\\&=&
\frac{1}{a_5}Y(t)
\frac{4^4h_n}{4N_cN_{x}}\sum_{n=1}^{N_l}
%\sum_{i=1}^{h_n}
\left(\frac{1}{2\cosh\frac{N_5}{2}a_5\wt{\lambda}(\lambda^{'(1)}_n)}\right)^2.
\end{eqnarray}
Here we use a relation for $i=j$
\begin{eqnarray}
&&
\frac{1}{h_n}f_n^{i,i}(t)=\frac{1}{h_{n,t}}\delta_{t\in{\cal S}_n},
\end{eqnarray}
where $h_{n,t}$ is a width of $t$ in ${\cal S}_n$.

(ii) Plane but rapidly oscillating eigenvector $\psi_n^{(i)}$ with
$C\simeq0$, which is typical for the excited mode with $i>1$.
In this case as was mentioned in the above a single summation
$\sum_{I}\psi_n^{(i)}(I)$ is suppressed and \eqn{eqn:propXeigen} has 
non-zero value only when $I=K$,
\begin{eqnarray}
&&
\frac{1}{h_n}\sum_{I}\sum_{K}
\psi_n^{(i)\dagger}(I)\VEV{I}{D_{N_5}^{-1}}{J}
\VEV{J}{\left(D_{N_5}^{-1}\right)^\dagger}{K}\psi_n^{(j)}(K)
\nn\\&&=
\frac{1}{h_n}\sum_{I}
\psi_n^{(i)\dagger}(I)\VEV{I}{D_{N_5}^{-1}}{J}
\VEV{J}{\left(D_{N_5}^{-1}\right)^\dagger}{I}\psi_n^{(j)}(I)
\nn\\&&=
\delta_{i,j}
 \VEV{I}{D_{N_5}^{-1}}{J}\VEV{J}{\left(D_{N_5}^{-1}\right)^\dagger}{I},
\end{eqnarray}
where $I,J$ unsummed.
$X_l$ then becomes 
\begin{eqnarray}
X_l(t) &=&
-\frac{a_5}{a^{10}}\sum_{n=1}^{N_l}\sum_{i>1}^{h_n}\frac{1}{h_n}f_n^{i,i}(t)
\left(\frac{1}{2\cosh\frac{N_5}{2}a_5\wt{\lambda}(\lambda^{'(i)}_n)}\right)^2
\sum_{b,\beta}
\VEV{I}{D_{N_5}^{-1}}{J}
\VEV{J}{\left(D_{N_5}^{-1}\right)^\dagger}{I}
\nn\\&=&
-\frac{a_5}{a^{10}}\frac{1}{4N_cn_{xyz}}
\sum_{n=1}^{N_l}\sum_{i>1}^{h_n}\frac{1}{h_{n,t}}\delta_{t\in{\cal S}_n}
\left(\frac{1}{2\cosh\frac{N_5}{2}a_5\wt{\lambda}(\lambda^{'(i)}_n)}\right)^2
\nn\\&&\times
\sum_{\vec{x}}\sum_{\alpha,a}
\sum_{b,\beta}
\VEV{\vec{x},t,\alpha,a}{D_{N_5}^{-1}}{y,\beta,b}
\VEV{y,\beta,b}{\left(D_{N_5}^{-1}\right)^\dagger}{\vec{x},t,\alpha,a},
\nn\\&=&
\frac{1}{a_5}Y(t)
\frac{1}{4N_cN_{x}}\sum_{n=1}^{N_l}\sum_{i>1}^{h_n}
\left(\frac{1}{2\cosh\frac{N_5}{2}a_5\wt{\lambda}(\lambda^{'(i)}_n)}\right)^2 .
\end{eqnarray}
In this case $m_{5q}$ is given by the same formula as in the previous
section with the completely localized eigenvector.
In addition we assume that the spin and color eigenfunction is always 
rapidly oscillating and set their enhance factor to be unity as is
discussed in the above.

The pion propagator in the denominator is expanded as
\begin{eqnarray}
Y(t) &=& \sum_{\vec{x}}\vev{P(x)P(y)}
\nn\\&=&
-\frac{a_5^2}{a^{10}}\sum_{\vec{x},a,\alpha}\sum_{b,\beta}
\Biggl(
\sum_{n,m=1}^{N_l}\sum_{i,j=1}^{h_n}
\frac{1}{h_n}\psi_n^{(i)}(I)\psi_m^{(j)}(I)
\VEV{n,i}{D_{N_5}^{-1}}{J}
\VEV{J}{(D_{N_5}^{-1})^\dagger}{m,j}
\nn\\&+&
\sum_{p,s,c}\sum_{k,s',d}
\frac{1}{N_p}U_\alpha(p,s)U_\alpha^\dagger(k,s')e^{i(p-k)x}
\VEV{p,s,a}{D_{N_5}^{-1}}{J}
\VEV{J}{(D_{N_5}^{-1})^\dagger}{k,s',b}
\Biggr).
\end{eqnarray}
By inserting a complete set $1=\sum\ket{I}\bra{I}$ the contribution from the
localized eigenvectors is rewritten in term of $Y(t)$.
\begin{eqnarray}
Y_l(t)&=&
-\frac{a_5^2}{a^{10}}\sum_{\vec{x},a,\alpha}
\sum_{n,m=1}^{n_l}\sum_{i,j=1}^{h_n}\frac{1}{h_n}\psi_n^{(i)}(I)\psi_m^{(j)}(I)
\sum_{b,\beta}
\sum_{z,\gamma,c}\braket{n,i}{z,\gamma,c}
\VEV{z,\gamma,c}{D_{N_5}^{-1}}{J}
\nn\\&&\times
\sum_{w,\delta,d}\VEV{J}{(D_{N_5}^{-1})^\dagger}{w,\delta,d}
\braket{w,\delta,d}{m,j}
\nn\\&=&
-\frac{a_5^2}{a^{10}}\sum_{\vec{x},a,\alpha}
\sum_{n,m=1}^{n_l}\sum_{i,j=1}^{h_n}\frac{1}{h_n}\psi_n^{(i)}(I)\psi_m^{(j)}(I)
\nn\\&&\times
\sum_{b,\beta}
\sum_{z,\gamma,c}
\frac{1}{h_n}\psi_n^{(i)\dagger}(z,\gamma,c)
\VEV{z,\gamma,c}{D_{N_5}^{-1}}{J}
\sum_{w,\delta,d}\VEV{J}{(D_{N_5}^{-1})^\dagger}{w,\delta,d}
\psi_m^{(j)}(w,\delta,d)
\nn\\&=&
-\frac{a_5^2}{a^{10}}\sum_{\vec{x},a,\alpha}
\sum_{b,\beta}
\VEV{I}{D_{N_5}^{-1}}{J}\VEV{J}{(D_{N_5}^{-1})^\dagger}
{I}_{I\in\cup_n{\cal S}_n}
\nn\\&=&
\frac{\sum_{n=1}^{N_l} h_n}{4N_cN_x}Y(t),
\end{eqnarray}
where we take average for $I\in\cup_n{\cal S}_n$.

The contribution from the plane-wave modes is the same as that of the
completely localized case by using the off diagonal property
$\VEV{p,s,a}{D_{N_5}^{-1}}{n,i}=0$.
Now we have a relation to write down $Y(t)$ in terms of $F(p)$ only,
\begin{eqnarray}
&&
Y(t\to\infty)=
-\frac{4N_cN_x}{4N_cN_x-hN_l}
\frac{a_5^2}{a^{10}}n_{xyz}'
\frac{4N_c}{N_p}\sum_{p}F(p),
\end{eqnarray}
where $ h N_l \equiv \sum_{n=1}^{N_l} h_n$,
and the anomalous quark mass $m_{5q}$ can be written as 
\begin{eqnarray}
m_{5q}&=&\frac{\lim_{t\to\infty}X(t)}{\lim_{t\to\infty}Y(t)}
\nn\\&=&
\frac{1}{a_5}\frac{1}{4N_cN_x}
\Biggl(
\sum_{n=1}^{N_l}4^4h_n
\left(\frac{1}{2\cosh\frac{N_5}{2}a_5\wt{\lambda}(\lambda^{'(1)}_n)}\right)^2
+\sum_{n=1}^{N_l}\sum_{i>1}^{h_n}
\left(\frac{1}{2\cosh\frac{N_5}{2}a_5\wt{\lambda}(\lambda^{'(i)}_n)}\right)^2
\nn\\&&
+4N_c\sum_{n}\rho_n\left(\frac{1}{2\cosh\frac{N_5}{2}a_5\wt{\lambda}_n}
\right)^2
\Biggr) .
\end{eqnarray}

In the actual case, the weight factor for the contribution from partially 
localized eigenstates lies between 1 and $4^4h_n$ depending on the index
$i$.
Therefore we adopt $\tilde h_n^{(i)}$ such that
 $1 \le \tilde h_n^{(i)} \le 4^4h_n$ as the weight factor in the text.

%%%%%%%%%%%%%%%%%%%%%%%%%%%%%%%%%%%%%%%%%%%%%%%%%%%%%%%%%%%%%%%%%%%%%%
\newcommand{\J}[4]{{#1} {#2}, #3 (#4)}
\newcommand{\AP}{Ann.~Phys.}
\newcommand{\CMP}{Commun.~Math.~Phys.}
\newcommand{\IJMP}{Int.~J.~Mod.~Phys.}
\newcommand{\MPL}{Mod.~Phys.~Lett.}
\newcommand{\NP}{Nucl.~Phys.}
\newcommand{\NPSup}{Nucl.~Phys.~B (Proc.~Suppl.)}
\newcommand{\PL}{Phys.~Lett.}
\newcommand{\PR}{Phys.~Rev.}
\newcommand{\PRL}{Phys.~Rev.~Lett.}
\newcommand{\PTP}{Prog. Theor. Phys.}
\newcommand{\Suppl}{Prog. Theor. Phys. Suppl.}
\newcommand{\RMP}{Rev. Mod. Phys.}
%%%%%%%%%%%%%%%%%%%%%%%%%%%%%%%%%%%%%%%%%%%%%%%%%%%%%%%%%%%%%%%%%%%%%%

%%%%%%%%%%%%%%%%% Figures %%%%%%%%%%%%%%%%%


%%%%%%%%%%%%%%%%% Figures %%%%%%%%%%%%%%%%%

\end{document}